\documentclass[reprint,superscriptaddress,amsmath,amssymb,amsfonts,aps,prl,floatfix,twocolumn]{revtex4-2}
\usepackage{times}
\usepackage{graphicx}
\usepackage{dcolumn}
\usepackage{float}
\usepackage{bm}
\usepackage{ulem}
\usepackage[colorlinks=true, citecolor=blue, linkcolor=blue, urlcolor=blue]{hyperref}
\usepackage{bbold}
\DeclareMathAlphabet\mathbfcal{OMS}{cmsy}{b}{n}
\usepackage{wasysym}
\usepackage{amssymb}
\usepackage{amsmath}
\usepackage[usenames,dvipsnames]{xcolor}

\begin{document}
\title{Quantum Geometric Kohn-Luttinger Superconductivity}
\author{Gal Shavit}
\affiliation{Department of Physics and Institute for Quantum Information and Matter, California Institute of Technology,
Pasadena, California 91125, USA}
\affiliation{Walter Burke Institute of Theoretical Physics, California Institute of Technology, Pasadena, California 91125, USA}
\author{Jason Alicea}
\affiliation{Department of Physics and Institute for Quantum Information and Matter, California Institute of Technology,
Pasadena, California 91125, USA}

\begin{abstract}

Coulomb repulsion can, counterintuitively, mediate Cooper pairing via the Kohn-Luttinger mechanism.  
However, it is commonly believed that observability of the effect requires special circumstances---e.g., vicinity of the Fermi level to van Hove singularities, significant lattice-induced band distortions, or non-trivial Fermi surface topologies.
Here we establish that quantum geometric properties of the constituent electrons can dramatically promote pairing from repulsion via dependence of screening on the quantum metric.  We demonstrate quantum-geometry-enhanced superconductivity in two microscopic models with tunable quantum geometry, highlighting the crucial roles of quantum metric anisotropy and inhomogeneity.  
Our analysis provides an experimentally accessible figure of merit for the importance of quantum geometry to inducing unconventional superconductivity, indicating its relevance to graphene multilayers.

\end{abstract}
\maketitle

{\bf \textit{Introduction.}}~The Kohn-Luttinger (KL) mechanism for superconductivity~\cite{KohnLuttingerPhysRevLett.15.524,KohnLuttinger2_PhysRev.150.202} describes how Coulomb interactions, screened by a Fermi liquid, may induce a superconducting instability. 
This effect was predicted to be rather weak for particles with parabolic dispersion~\cite{KohnLuttingerPhysRevLett.15.524,lowdensityKLPhysRevLett.20.187,chubukov_parabolic_kagan1988possibility}, and potentially even weaker in 2D systems, where effective attraction arises at higher-order in perturbation theory~\cite{chubukov_2deg}.
Various factors can, however, greatly promote KL superconductivity:
distortion of the electronic dispersion by an underlying lattice~\cite{Kivelson_Hubbard_KL_PhysRevB.81.224505}, the presence of Fermi surfaces of different sizes~\cite{kivelson_raghu_2degwithbandsPhysRevB.83.094518,chubukov_kivelson_2component2deg}, multiple Fermi pockets~\cite{chubukov_pnictidesPhysRevB.82.214515,Cubukov_review}, van Hove singularities~\cite{graphen_vhs_klPhysRevB.78.205431,chubukov_grapheneNandkishore2012}, and annular Fermi surfaces~\cite{annular_PhysRevLett.127.247001}.
More recently, this mechanism was proposed to play a role in superconducting graphene multilayers, which host an amalgam of the above properties~\cite{KL_TBG,PacoTBGdoi:10.1073/pnas.2107874118,Paco_RTG_PhysRevB.105.075432,AlejandroPacoBLG,KWANBLG}.

\begin{figure}
    \centering
    \includegraphics[width=7.5cm]{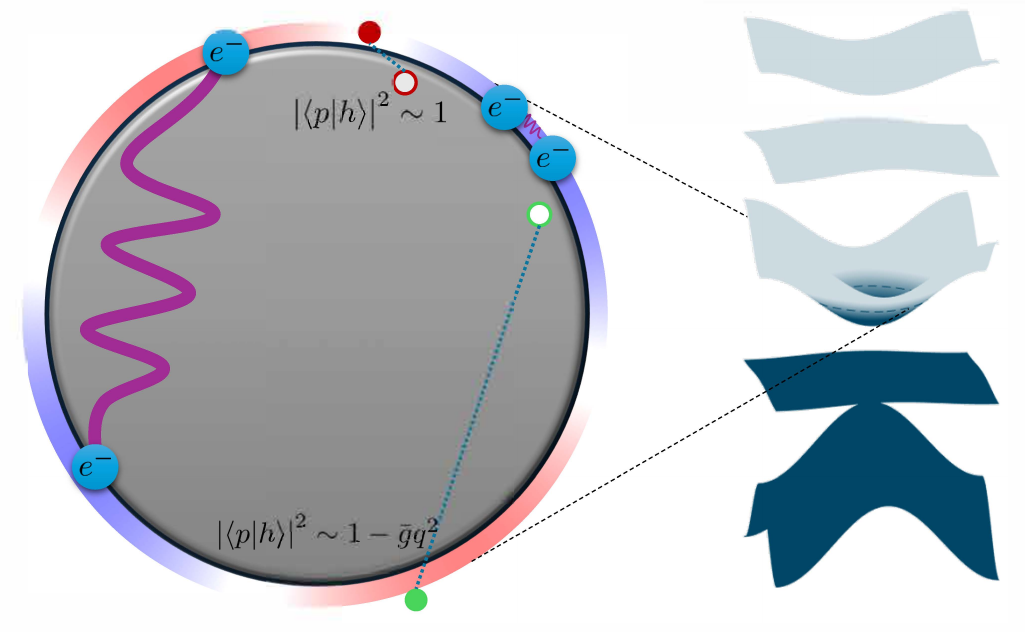}
    \caption{{\bf Mechanism underlying quantum-geometry-enhanced superconductivity.}  
    Right: Multiband dispersion of a 2D material with partially filled bands.
    Left:
    Effective Fermi surface interactions.
    Small-momentum interactions (thin purple line) are efficiently screened by small momentum particle-hole (red circles), whose wavefunction overlap is close to unity.
    Large-momentum interactions (thick purple line), by contrast, are poorly screened since 
    large-momentum particle-hole excitations (green circles) exhibit significantly lower wavefunction overlap.
    The overlap reduction is approximately quadratic in momentum, with a coefficient $\bar{g}$ corresponding to the Fermi-surface-averaged quantum metric (see text).
    An effective interaction that increases with momentum promotes a sign-changing superconducting order parameter that minimizes the interaction energy (red and blue halo).}
    \label{fig:schematics}
\end{figure}

In this Letter, we demonstrate the emergence of KL superconductivity \textit{driven by non-trivial quantum geometry of the interacting electron wavefunctions}.   Quantum geometry in momentum space~\cite{Riemannian_quantum_geometryProvost1980,vanderbilt2018berry} has emerged as a fascinating field of study and influences various solid-state phenomena~\cite{Vanderbiltmaximmalylocalized,Martin_geometry_polariaztionPhysRevB.62.1666,Pgeometry_magnetic_susceptihysRevB.91.214405,PESINCME_PhysRevB.92.235205,geometry_bounds_holder_mitscherlingPhysRevB.105.085154,dassarma_geometry_goldstonePhysRevB.102.165118,Kaplan_non_linear_PhysRevLett.132.026301,Capacitance_holder_Komissarov2024,geometrymicrowaveSC_arora2024quantumgeometryinducedmicrowave} including superfluid stiffness in flat-band superconductors~\cite{superfluid_og_Peotta2015,superfluid_tbgPhysRevLett.123.237002,superfluid_review_Törmä2022,superfluid_bounds_herzog_peri,ktlaw_gl_metricPhysRevLett.132.026002}
as well as the stability of the fractional Chern insulators~\cite{FCIreviewNeupert_2015,RoytraceconditionFCI,BergholtzLatticeConstantBerry,FCI_Khalef_moire,FCI_TBG_parker2021fieldtuned,simonrudnerindicators,Jackson2015Roynumerical,BergholtzLatticeConstantBerry,FCI_TBG_parker2021fieldtuned,numerical_tmd_pressure_PhysRevResearch.5.L032022,TMD_numerical_Devakul,TMD_numerical_wang2023fractional,shavit2024FCI}.  
Novel geometry-driven superconducting phenomena, including spin-fluctuation-mediated pairing~\cite{QuantumGeometryFerrofluctuationsPhysRevLett.132.036001} and anomalous Fulde-Ferrell-Larkin-Ovchinnikov states~\cite{additional1,additional2,additional3,additional4,FFLOgeometric_sun2024flatbandfflostatequantum} have also been recently reported.

The relation between quantum geometry and KL superconductivity admits an intuitive physical picture sketched in Fig.~\ref{fig:schematics}: screening by virtual particle-hole excitations is heavily affected by their quantum distance (or wavefunction overlap), endowing further structure to the momentum-dependent pairing vertex.  
We show that this additional structure can enhance the critical temperature by orders of magnitude (compared to `pure' KL superconductivity) while also fundamentally altering the dominant pairing channel.

Our analysis reveals that the average `Fubini-Study metric' around the Fermi surface provides a key figure of merit for the importance of quantum geometric effects on pairing instabilities.  
As one immediate implication, a quantum metric that is inhomogeneous in momentum space naturally yields non-trivial dependence of the superconducting coupling constant on density.
Superconducting materials with highly nontrivial quantum geometry in turn may exhibit critical temperatures that vary as a function of density more dramatically than is to be expected from the corresponding variation in the density-of-states (DOS).
Enhancement of unconventional pairing by quantum geometry is potentially key to understanding the underlying superconducting mechanism in recent experimental discoveries in geometrically rich quantum materials, where peculiar trends in the density dependence of superconductivity are ubiquitously observed~\cite{ZhouYoungBLGZeeman,RTGsuperconductivityZhou2021,nadj_ISOC_BBGBLGZhang2023,TDBG_folk_Su2023,Quasicrystal_Uri2023,BBG_electronside_li2024tunable,LongJuyang2024diverseimpactsspinorbitcoupling,Nadj_zhang2024twistprogrammablesuperconductivityspinorbitcoupled,young_nadj_BBG_RTG_SC,Young_patterson2024superconductivityspincantingspinorbit,Young_choi2024electricfieldcontrolsuperconductivity,WSE2_cornellxia2024unconventionalsuperconductivitytwistedbilayer,Wse2_hysteresis_CoryDean_guo2024superconductivity}.
Furthermore, the principles highlighted here help guide the discovery of new superconducting materials, as well as unconventional superconductivity in previously unexplored regimes.

{\bf \textit{General Theory.}}~Consider an interacting 2D system whose Hamiltonian is projected onto a set of $N_f$ active bands, labeled by `flavors' $\alpha$, that cross the Fermi energy.  At a given momentum $\bf k$, the projection operator onto a given active band reads $P_{\bf k}^\alpha=|{\mathsf u}_{\bf k}^\alpha\rangle\langle {\mathsf u}_{\bf k}^\alpha|$, where $|{\mathsf u}_{\bf k}^\alpha\rangle$ are the associated Bloch wavefunctions.
Evolution of these wavefunctions with $\bf k$ generates emergent quantum geometry.
The projected Hamiltonian is
\begin{equation}
    H = \sum_{{\bf k},\alpha} \left(\epsilon_{{\bf k},\alpha}-\mu\right)\psi^\dagger_{{\bf k},\alpha}\psi_{{\bf k},\alpha}
    +
    \frac{1}{2A}\sum_{\bf q} V_{\bf q}\Tilde{\rho}_{\bf q}\Tilde{\rho}_{\bf -q}, \label{eq:projectedBareHamiltonian}
\end{equation}
where $\psi_{{\bf k},\alpha}$ are fermionic annihilation operators, $\mu$ is the chemical potential, $\epsilon_{{\bf k},\alpha}$ are band energies, $V_{\bf q}=\frac{2\pi e^2}{\epsilon}\frac{\tanh{qd}}{q}$ is the screened Coulomb interaction ($q=\left|{\bf q}\right|$, and $d$ is the distance to screening metallic gates), and $A$ is the system's area.
The projected density operator is
$\Tilde{\rho}_{\bf q}=\sum_\alpha \Lambda^\alpha_{\bf k,k+q}\psi^\dagger_{{\bf k+q},\alpha}\psi_{{\bf k},\alpha}$, with $\left|\Lambda^\alpha_{\bf k,k'}\right|^2= {\rm Tr}\left[P_{\bf k}^\alpha P_{\bf k'}^\alpha\right]$.
Hereafter we assume that $H$ preserves time-reversal symmetry; the $N_f$ flavors then come in pairs, such that each flavor $\alpha$ admits a time-reversed partner $\bar{\alpha}$, which together satisfy
$\epsilon_{{\bf k},\alpha}=\epsilon_{-{\bf k},\bar{\alpha}}\equiv \epsilon_{\bf k}$ and
$\Lambda^\alpha_{\bf k,k'}=\left(\Lambda^{\bar{\alpha}}_{\bf k,k'}\right)^*\equiv\Lambda_{\bf k,k'}$.

The repulsion $V_{\bf q}$ is effectively screened by electron-hole fluctuations of the Fermi liquid around the Fermi surface.
In the random phase approximation (RPA), the screened interaction is 
\begin{equation}
    V_{\bf q}^{\rm RPA}=\frac{V_{\bf q}}{1+\Pi_{\bf q}  V_{\bf q}},
\end{equation}
where 
\begin{equation}
    \Pi_{\bf q}= - N_f \sum_{\bf k}
    \left|\Lambda_{\bf k,k+q}\right|^2
    \frac{ n\left(\xi_{\bf k+q}\right) - n\left(\xi_{\bf k}\right) }
    {\xi_{\bf k+q} - \xi_{\bf k}}
    \label{eq:staticpol}
\end{equation}
is the static polarization ($\xi_{\bf k}=\epsilon_{\bf k}-\mu$ and $n\left(x\right)=\left(1+e^{x/T}\right)^{-1}$ is the Fermi-Dirac distribution with temperature $T$; we henceforth set $T=0$ in $\Pi_{\bf q}$ calculations).  Note in particular the dependence on quantum geometry through the $\Lambda$ factor above.  To proceed we specialize to rotationally-symmetric models with 
parabolic dispersion $\epsilon_{\bf k}=k^2/\left(2m\right)$ ($k=\left|{\bf k}\right|$ and $m$ denotes the effective electron mass), allowing us to isolate quantum geometrical effects from other superconductivity-enhancing mechanisms mentioned earlier.
Rotation symmetry allows us to simplify form-factors appearing in Eq.~\eqref{eq:staticpol} via
$\left| \Lambda_{\bf k+q,k}\right|^2 \to {\cal F}\left(q,k,\beta\right)$
with $\beta$ the angle between $\bf k$ and $\bf q$.

In the absence of quantum geometrical effects, all $\Lambda_{\bf k,k'}=1$, and the static polarization for any $q<2k_F$ is constant, $\Pi_0=N_f\frac{m}{2\pi}$~\footnote{For parabolic dispersion in two dimensions with trivial quantum geometry, the analytic expression of the static polarization is $\Pi_{\bf q}=\Pi_0
\left(1-
{\rm Re}\sqrt{1-\left(\frac{2k_F}{q}\right)^2}
\right)$.}.
Nontrivial band geometry, by contrast, engenders momentum dependence at momenta much smaller than $2k_F$, where one finds
\begin{equation}
    \Pi_{\bf q} \approx \Pi_0
    \left[
    1-q_\mu q_\nu \left\langle{g}_{\mu\nu}\right\rangle_{\rm FS}
    \right] + O\left(q^3\right).\label{eq:underscreenedPi}
\end{equation}
Here $\left\langle\cdot\right\rangle_{\rm FS}$ denotes Fermi-surface averaging, and 
$g_{\mu \nu} = {\rm Tr}\left[\partial_\mu P_{\bf k} \partial_\nu P_{\bf k}\right]$ is the Fubini-Study metric.
The geometry-induced momentum-dependent polarization correction is negative---thus suppressing screening at higher momentum. 
In particular, while the bare $V_{\bf q}$ monotonically decreases with $q$, the screened interaction may obtain a positive curvature at small momentum~\cite{QuantumGeometryFerrofluctuationsPhysRevLett.132.036001,SM}: 
\begin{equation}
    \partial_{q}^{2}V_{{\bf q}}^{{\rm RPA}}|_{q\to0}
    \propto
    -\left(1-\frac{3N_{f}\bar{g}}{\lambda_{{\rm TF}}d}\right),\label{eq:underscreen}
\end{equation}
where $\bar{g}\equiv \left\langle{{\rm tr}g}\right\rangle_{\rm FS}$ and $\lambda_{{\rm TF}}=\epsilon/\left(e^{2}m\right)$ is the Thomas-Fermi screening length.
If
$\bar{g}>\frac{\lambda_{TF}d}{3N_f}$,
the repulsion grows with momentum separation.
Such behavior of the repulsive interaction naturally promotes superconductivity with an order parameter that changes sign along the Fermi surface (Fig.~\ref{fig:schematics}).
Eq.~\eqref{eq:underscreen} is analogous to (yet distinct from) the susceptibility-curvature condition discussed in Ref.~\cite{QuantumGeometryFerrofluctuationsPhysRevLett.132.036001}.
Notice that, given Eq.~\eqref{eq:underscreenedPi}, the curvature of $V_{\bf q}^{\rm RPA}$ is always positive for Hubbard-like bare interactions, i.e., when $V_{\bf q}$ becomes momentum independent. 

Our analysis thus provides a rule of thumb---which we dub the \textit{geometric underscreening criterion}---for the importance of quantum geometric effects in the KL mechanism: 
If the length scale $\ell_{\rm FS}\equiv \sqrt{\bar g}$ is comparable to or exceeds the geometric mean of $d$ and $\lambda_{\rm TF}$, then quantum geometric effects are relevant due to inefficient screening of large-momentum interactions.
We note that ${\rm tr} g$, and by extension $\ell_{\rm FS}$, are related to the minimal width of Wannier functions for this band~\cite{Vanderbiltmaximmalylocalized}.

Using the screened interaction, and assuming pairing between time-reversed flavors, we solve the self-consistent Bardeen-Cooper-Schreiffer (BCS) gap equation,
\begin{equation}
    \Delta_{\bf k}=
    -\int\frac{d{\bf k'}}{\left(2\pi\right)^2}
    u_{\bf kk'}
    \frac{\tanh{\frac{E_{\bf k'}}{2T}}}{2E_{\bf k'}}
    \Delta_{\bf k'}.\label{eq:fullBCS}
\end{equation}
Here, the Bogoliubov spectrum is $E_{\bf k}=\sqrt{\xi_{\bf k}^2+\left|\Delta_{\bf k}\right|^2}$, and the geometry-dependent interaction matrix is
$u_{\bf kk'}=\left|\Lambda_{\bf k,k'}\right|^2V_{\bf k-k'}^{\rm RPA}$.
To find the critical temperature $T_c$, we linearize Eq.~\eqref{eq:fullBCS}, assume negligible radial dependence of the integrand (justified by overwhelmingly near-Fermi-surface contribution to the integral), and utilize the parabolic spectrum.
These simplifications yield~\cite{SM} 
\begin{equation}
    \Delta_{\theta} = -\ln\left(\frac{W}{T_c}\right)\frac{m}{2\pi}\int\frac{d\theta'}{2\pi}{u}^{k_F}_{\left(\theta-\theta'\right)}\Delta_{\theta'},\label{eq:BCSmain}
\end{equation}
where $W\sim E_F$ is an energy cutoff, and the  interaction ${u}^{k_F}_{\left(\theta\right)}$ corresponds to $u_{\bf kk'}$ connecting points with Fermi momentum $k_F$ separated by an angle $\theta$.

\begin{figure*}
    \centering
    \includegraphics[width=18cm]{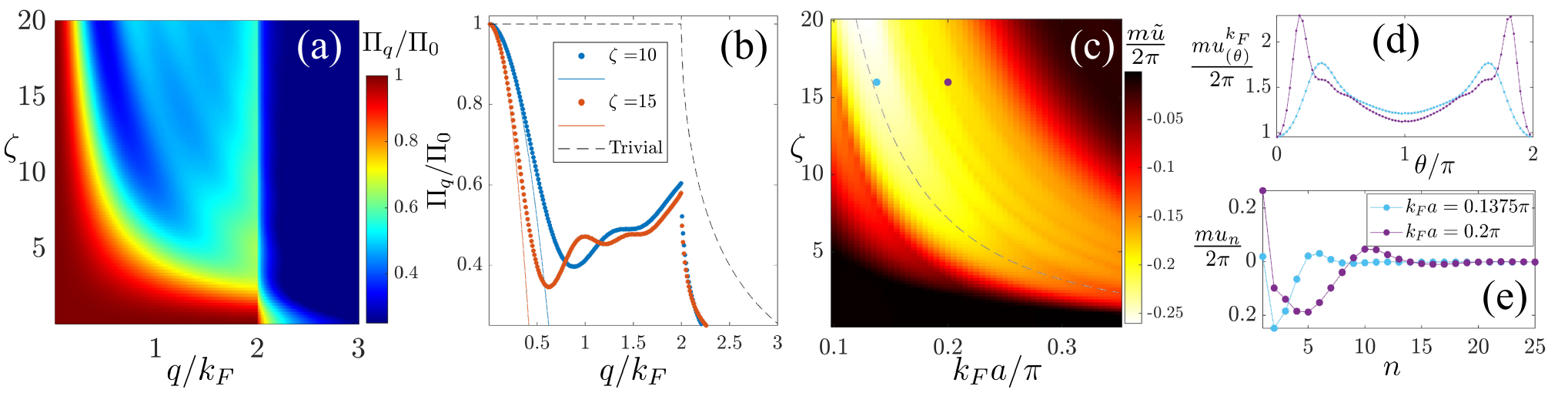}
    \caption{
    {\bf Results for the TM model [Eq.~\eqref{eq:MAINtmHamiltonian}].}
    (a)
    Static polarization as function of $\zeta$,
    with $k_F=0.2\pi/a$.
    (b)
    Cuts through (a) at fixed $\zeta$ (see legend), compared to the case of trivial quantum geometry (dashed black line).
    Solid lines correspond to the small-momentum expansion $\Pi_q\propto 1-\bar{g}q^2$.
    (c)
    Optimal coupling constant as a function of $\zeta$ and Fermi momentum assuming $d/a=5$ and $\lambda_{\rm TF}/a=0.25$.
    Dashed line corresponds to
    $\left(k_F a \right)^2=\frac{\sqrt{8}}{\zeta}$,
    reflecting optimal conditions for $d$-wave pairing ($\tilde n=2$).
    (d)
    Interaction kernel ${u}^{k_F}_{\left(\theta\right)}$ [Eq.~\eqref{eq:BCSmain}] and (e) angular decomposition corresponding to the colored points in (c).}
    \label{fig:TMresults}
\end{figure*}

The leading superconducting instability is obtained by utilizing rotational symmetry and performing an angular decomposition ${u}^{k_F}_{\left(\theta\right)}=u_0/2 +\sum_{n>0} u_n \cos\left(n\theta\right) $.
Notice that if ${u}^{k_F}_{\left(\theta\right)}$ has positive curvature at $\theta=0$, $\partial^2_\theta {u}^{k_F}|_{\theta=0}<0$, as implied by fulfilling the geometric underscreening criterion [Eq.~\eqref{eq:underscreen} and following discussion], at least one $u_n$ coefficient is necessarily negative.
Denoting by $\tilde{u}$ the maximum-amplitude negative coefficient in this expansion, and its order by $\tilde{n}$, one obtains
$T_c\approx W\exp\left(-\frac{2\pi}{m\left|\tilde{u}\right|}\right)$, with a gap function $\propto \cos\left(\tilde{n}\theta\right)$.  Crucially, the geometric mechanism drives more attractive $u_n$ coefficients at lower $n$, facilitating vastly larger $T_c$'s.  
At this level of analysis, due to the rotational symmetry all linear combinations of $\exp\left(\pm i\tilde{n}\theta\right)$ are degenerate solutions of the gap equation.
In the Supplementary Material (SM), we relax the time-reversal-symmetric pairing assumption, demonstrating one of many unexplored possibilities the geometric KL mechanism unlocks, namely, promoting chiral superconducting instabilities~\cite{SM}. 

{\bf \textit{Analysis.}}~We now examine two concrete models exemplifying the power of quantum geometry in promoting superconductivity. 
First, we consider a `tunable metric' (TM) lattice model (related to Refs.~\cite{tunablemetricHofmann2022,tunableMetricPRL_PhysRevLett.130.226001}) with single-particle Hamiltonian, prior to projection onto the active bands, expressed as
\begin{equation}
    H_{\rm TM} = H_{\zeta} + H_{\rm n.n.}.
    \label{eq:MAINtmHamiltonian}
\end{equation}
The first term,
\begin{equation}
    H_{\zeta}=\frac{E_g}{2}\sum_{{\bf k}}c_{{\bf k}}^{\dagger}\left[\sin\left(\zeta{\alpha_{\bf k}}\right)\tau_{x}+\sigma_{z}\cos\left(\zeta{\ensuremath{\alpha_{{\bf k}}}}\right)\tau_{y}\right]c_{{\bf k}},
\end{equation}
manifests two flat bands separated by an energy gap $E_g$.
Here $c_{\bf k}$ is a spinor describing fermions hopping on a square lattice with momentum $\bf k$.   The orbital degrees of freedom are addressed by Pauli matrices $\tau_i$, time-reversed flavors are addressed by $\sigma_i$, and ${\alpha_{{\bf k}}}=\left[\cos\left(k_{x}a\right)+\cos\left(k_{y}a\right)\right]$, with lattice constant $a$.
The second term, $H_{\rm n.n.}$, encodes intra-orbital intra-flavor nearest-neighbor hopping with magnitude $t\ll E_g$.
It endows identical dispersion to the two bands, without altering their corresponding wavefunctions.
When the Fermi level resides near either the top or bottom of a band, the spectrum is approximately parabolic, with effective mass $m\approx\left(a^2 t\right)^{-1}$.
Without loss of generality, we restrict to the bottom of the valence band.
The resulting parabolic model exhibits widely tunable quantum-geometry dictated by $\zeta$.
We note that the TM model possesses a unique quantum-geometrical property, namely form-factors between Fermi surface wavefunctions are identically one.
This property is not limited to the small-momentum regime explored here, but rather descends from the Hamiltonian~\eqref{eq:MAINtmHamiltonian} itself.
Consequently, for this model $u_{\bf kk'}=V_{\bf k-k'}^{\rm RPA}$.

Figure~\ref{fig:TMresults}(a) displays the $\zeta$ dependence of the static polarization.  
At small momenta, it decays quadratically as $1-\bar{g}q^2$, with $\bar{g}= \frac{\left(\zeta a\right)^2}{8} \left(k_F a\right)^2$  as discussed above; see Fig.~\ref{fig:TMresults}(b).
Oscillations at larger momenta can be traced to the oscillatory form factors
${\cal F}\left(q,k,\beta\right)=  \cos^{2}\left[\frac{\zeta}{4}\left(2kq\cos\beta+q^{2}\right)\right]$.
In Fig.~\ref{fig:TMresults}(c) we plot the BCS coupling constant as a function of $\zeta$ and the Fermi momentum (assuming the same interaction as in Eq.~\eqref{eq:projectedBareHamiltonian}).
Once $\zeta$ is sufficiently large that quantum metric effects become significant, superconductivity becomes appreciable. 
For reference, quantum geometrical effects disappear in the limit $\zeta\rightarrow 0$.
Using the same parameters as in Fig.~\ref{fig:TMresults}, here we find coupling constants on the order of $10^{-3}$---two orders of magnitude smaller compared to the geometrically-enriched scenario.
Considering the exponential dependence of $T_c$ on the coupling strength, superconductivity is virtually absent in this limit, but becomes colossally enhanced by quantum geometry.

Notably, for given quantum geometry, i.e., fixed $\zeta$, Fig.~\ref{fig:TMresults}(c) reveals an optimal electron density for superconductivity; see dashed line.
This trend follows from the static polarization's momentum dependence.  
At small $q$, $\Pi_q$ exhibits a downward decline dictated by $\bar{g}$, while 
at intermediate momenta, comparable to $k_F$, the function oscillates around $\sim \Pi_0/2$.
The crossover occurs at roughly $q_{*}a=\frac{1}{\frac{\zeta}{2}k_{F}a}$, where the RPA-screened interaction is expected to be the strongest.
Thus, for a given order $\tilde n$---i.e., superconducting gap symmetry---one expects a corresponding optimal Fermi momentum, minimizing the interaction energy of the sign-changing pair function; see Fig.~\ref{fig:schematics}.
In our analysis the strongest superconductivity appears in the $\tilde{n}=2$ ($d$-wave) channel, optimized along the line $\left(k_F a \right)^2=\frac{\sqrt{8}}{\zeta}$ (corresponding to $q^*=k_F/\sqrt{2}$).

Figures~\ref{fig:TMresults}(d,e) further explore the interaction near optimal density (teal) in comparison to the `overdoped' regime (purple).  
Although $\bar{g}$ is larger in the latter case, the interaction also exhibits sharper angular dependence that spreads out the interaction angular over a broader range; in turn, the optimal coefficient is eventually less attractive. 

\begin{figure}
    \centering
    \includegraphics[width=8.8cm]{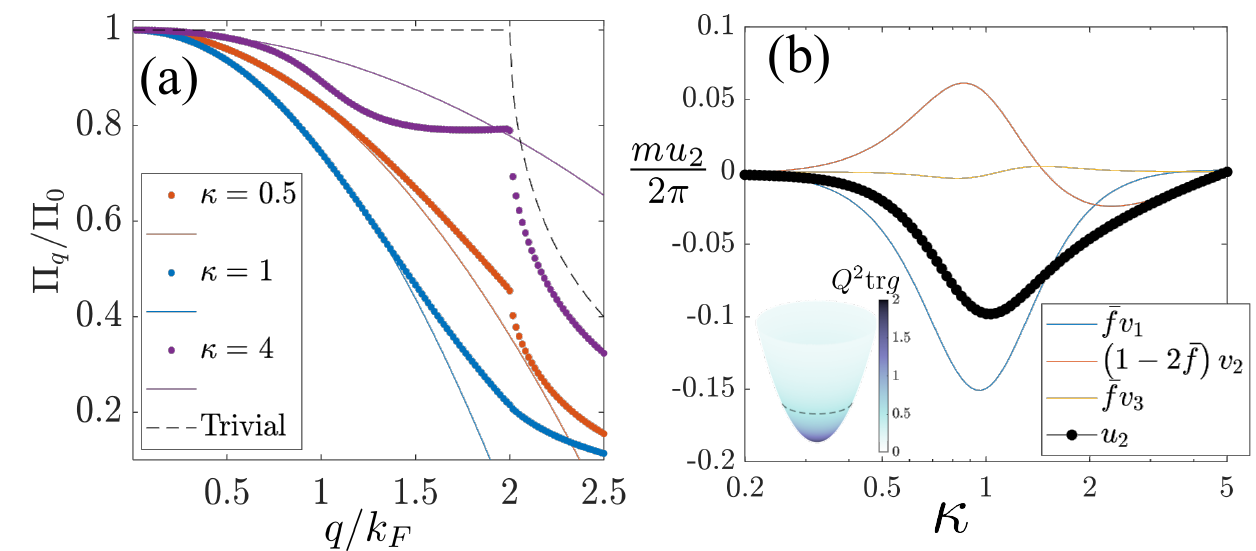}
    \caption{
    {\bf Results for the BHZ model [Eq.~\eqref{eq:TBImainH}].}
    (a)
    Static polarization at different $\kappa =  k_F/Q$ (dots).
    Solid lines represent the low-momentum expansion
    $\Pi_q\propto 1- \left(q/k_F\right)^2 \left[\kappa/\left(\kappa^2+1\right)\right]^2$.
    (b)
    Leading coupling constant $u_2$ (black) and its decomposition to components inherited from the RPA interaction (see text).
    Here, $Qd=1.5$ and $Q\lambda_{\rm TF}=0.08$.
    Inset: Distribution of quantum metric along the parabolic band in the BHZ model. Dashed line marks $\kappa=1$.}
    \label{fig:TBIresults}
\end{figure}

Next, we explore geometry-induced superconducting instabilities in the more familiar Bernevig-Hughes-Zhang (BHZ) model that features topological band inversion~\cite{QSHbernevig}.
The unprojected single-particle Hamiltonian reads
\begin{equation}
    H_{\rm BHZ}=\sum_{{\bf k}}\Psi_{{\bf k}}^{\dagger}\left[\
    vk_x\sigma_z\tau_x
    +vk_y\tau_y
    +\left(\frac{k^2}{2m}+M\right)\tau_z
    \right]\Psi_{{\bf k}}\label{eq:TBImainH}
\end{equation}
with $\Psi_{\bf k}$ a fermionic spinor living in the Hilbert space spanned by Pauli matrices $\tau_i$, $\sigma_i$.  
Judiciously setting $M=-mv^2/2$ yields the parabolic spectrum
$\epsilon_{\bf k}=\pm\frac{k^2+Q^2}{2m}$,
where $Q=mv$, and a corresponding quantum metric $g_{\mu\nu}=\delta_{\mu\nu}\frac{Q^2}{\left(k^2+Q^2\right)^2}$~\cite{SM}. 
The metric is concentrated around the band bottom (see inset of Fig.~\ref{fig:TBIresults}(b)) and 
saturates the so-called trace condition, i.e., ${\rm tr}g=\left|{\cal B}\right|$ at all $\bf k$, with $\cal B$ the Berry curvature~\cite{trithep_tan2024parentberrycurvatureideal}.

The low-momenta static polarization behaves as 
$\Pi_{q}\propto 1-\left(q/k_F\right)^2\bar{f}$,
where
$\bar{f}=k_F^2\bar{g}=\frac{\kappa^2}{\left(\kappa^2+1\right)^2}$,
and
$\kappa=k_F/Q$.
Figure~\ref{fig:TBIresults}(a) presents $\Pi_q$ for several $\kappa$ values;
evidently, maximal suppression of $\Pi_q$ along the Fermi surface is achieved for $\kappa\approx 1$.
Contrary to the TM model, the form-factors ${\cal F}\left(q,k,\beta\right)$ contain no oscillatory dependence on $q$. 
Thus there is no transitional momentum scale for this model, and
$\bar f$ is an appropriately quantifies the effects of quantum geometry within the RPA treatment.

The quantity $\bar{f}$ appears also in the form factors relating wavefunctions on the Fermi surface at relative angle $\theta$: 
${\cal F}^{k_{F}}\left(\theta\right)=\left(1-2\bar{f}\right) + 2\bar{f}\cos\theta$.
Decomposing the interaction,
${u}^{k_F}_{\left(\theta\right)}={\cal F}^{k_{F}}V^{{\rm RPA},k_F}$, 
we extract the angular harmonics 
\begin{equation}
    u_{n}=\left(1-2\bar{f}\right)v_{n}+\bar{f}\left(v_{\left|n-1\right|}+v_{n+1}\right),\label{eq:constructivedecompositionTBI}
\end{equation}
where the $v_i$ are obtained from
$V^{{\rm RPA},k_F}=v_{0}/2+\sum_{n>0}v_{n}\cos\left(n\theta\right)$.
The characteristic shape of $V^{{\rm RPA},k_F}$ in the BHZ model is dominated by two coefficients: the constant repulsion $v_0>0$ and the $p$-wave-like attraction $v_1<0$.  Quantum geometric effects generate an additional $\cos\theta$ term in the form factors $\mathcal{F}^{k_F}$ that, in conjunction with the attractive $v_1$, yield a dominant superconducting instability in the $d$-wave channel.  That is, the dominant negative coefficient of the interaction kernel in~\eqref{eq:BCSmain} is $u_2\approx -\bar{f}\left|v_1\right|$.
As $\bar{f}$ is maximized around $\kappa=1$, one concludes again that this regime is optimal for superconductivity.
Figure~\ref{fig:TBIresults}b illustrates the $\kappa$ dependence of $u_2$, along with its decomposition according to Eq.~\eqref{eq:constructivedecompositionTBI}---confirming maximal attraction near $\kappa = 1$ as well as the dominant contribution from $v_1$.  

As an aside, quantum geometric effects described here may profoundly impact the $u_0$ component as well.
While $u_0$ remains positive (repulsive), not leading to superconductivity on its own, one finds $u_{0}=\left(1-2\bar{f}\right)v_{0}-2\bar{f}\left|v_1\right|$.
Hence, when $\bar{f}$ is maximized the $s$-wave Coulomb repulsion is minimized.
In the presence of retarded pairing interactions~\cite{AndersonMorel,tolmachev1962logarithmic} (e.g., generated by a non-KL mechanism), this effect may further promote superconductivity and engender pronounced electron-density dependence of $T_c$ 
beyond that arising from DOS variation~\cite{SM}. 

{\bf \textit{Discussion.}}~The physical mechanism at work, driving a Fermi liquid with non-trivial geometrical properties into a superconductor, is form-factor-driven underscreening.
At non-zero momenta, particle-hole excitations across a ``geometrically-charged'' Fermi surface with non-negligible $\bar{g}$ are increasingly suppressed.
The resultant distortion of the interaction on the Fermi surface generically promotes angular harmonics with effective attraction, and eventually a corresponding Cooper instability.

One might expect that the geometry-derived Fermi surface form factors ${\cal F}^{k_{F}}$ should counteract the underscreening and weaken or eliminate the superconducting instabilities.
We demonstrated that this is not necessarily the case, and now argue that these geometric effects do not \textit{generically} conflict. 
The geometric underscreening is contributed by form-factors between the Fermi sea and the complimentary unoccupied states.
Conversely, ${\cal F}^{k_{F}}$ is determined solely by the quantum distance \textit{along the Fermi surface}.

Borrowing General Relativity nomenclature, an anisotropic quantum metric may completely decouple these two effects.
The TM model is a perfect example, possessing a metric which is purely radial, implying zero quantum distance between points on the Fermi surface.
Conversely, if the metric is isotropic yet inhomogenous, as in the BHZ model, the on-Fermi-surface effect can be secondary in nature to the Fermi-sea contribution.
The latter can benefit more efficiently from areas of concentrated quantum metric.
Finally, in the non-generic case of an isotropic and homogeneous metric, the phenomenon we describe may indeed be suppressed, as demonstrated in the SM for a Landau level-like band~\cite{SM}.

The models discussed above all have benign parabolic dispersion and featureless DOS.
Nevertheless, the superconducting coupling strength depends strongly on the electron density; see Figs.~\ref{fig:TMresults}(c) and \ref{fig:TBIresults}(b).
The uncovered density sensitivity, which in the cases we explore is clearly DOS-independent, may shed some light on the non-trivial superconducting $T_c$ dependence in recent experiments~\cite{ZhouYoungBLGZeeman,RTGsuperconductivityZhou2021,nadj_ISOC_BBGBLGZhang2023,TDBG_folk_Su2023,Quasicrystal_Uri2023,BBG_electronside_li2024tunable,LongJuyang2024diverseimpactsspinorbitcoupling,Nadj_zhang2024twistprogrammablesuperconductivityspinorbitcoupled,young_nadj_BBG_RTG_SC,Young_patterson2024superconductivityspincantingspinorbit,Young_choi2024electricfieldcontrolsuperconductivity,WSE2_cornellxia2024unconventionalsuperconductivitytwistedbilayer,Wse2_hysteresis_CoryDean_guo2024superconductivity}.
These experiments ubiquitously observe superconducting regions where the critical temperature varies significantly more quickly than the exponential dependence on the DOS would imply. 
Notably, these materials possess non-trivial quantum geometric properties, with the quantum metric having considerable magnitude near the relevant experimentally accessible Fermi surfaces.

This result highlights another salient aspect of this work.
Namely, cementing the quantum geometrical properties of the underlying Fermi liquid, e.g., the Fermi-surface quantum metric $\bar{g}$, as a fundamental property with potentially far-reaching implications. 
Furthermore, viewing the present superconductivity scenario as an instability of a geometrically-charged Fermi liquid, our work paves the way for exploration of other intriguing consequences for such a parent phase.
Possibilities include density-wave and Pomeranchuk instabilities, and collective excitation phenomena (e.g., magnons and plasmons), all of which
may display rich behavior when the quantum metric is substantial near the Fermi surface.  

The physics described here is relevant to the recently expanding field of superconductivity in rhombohedral graphene.
A ubiquitous feature in those systems is the presence of small trigonal-warping-induced Fermi pockets in the normal state~\cite{ZhouYoungBLGZeeman,RTGsuperconductivityZhou2021,nadj_ISOC_BBGBLGZhang2023,BBG_electronside_li2024tunable,LongJuyang2024diverseimpactsspinorbitcoupling,Nadj_zhang2024twistprogrammablesuperconductivityspinorbitcoupled,young_nadj_BBG_RTG_SC,Young_patterson2024superconductivityspincantingspinorbit,Young_choi2024electricfieldcontrolsuperconductivity}.
Intriguingly, the Fubini-Study metric of the relevant bands is mostly concentrated in the vicinity of these pockets.
Whereas the magnitude of the metric is $\sim 30$ nm$^2$ in these regions, the Thomas-Fermi wavelength is $\lambda_{\rm TF}\sim 1$ nm in the relevant regimes~\cite{DasSarmaBLGphonons,StonerBlockadeBBGPhysRevB.108.024510}.
Given that the gate distance in these experiments is on the order of $\sim10$ nm, the criterion implied by Eq.~\eqref{eq:underscreen} can easily be satisfied.
These simple estimates highlight relevance of quantum geometry to the observed superconductivity, an effect that previously received little attention in this context~\footnote{For example, Ref.~\cite{KWANBLG} which considered KL superconductivity in the trigonal-warping pockets of bilayer graphene, explicitly suppressed all form-factor dependence.
Ref.~\cite{LiangFugeier2024chiraltopologicalsuperconductivityisospin} employed a single-band limit for extracting the RPA screened interaction in rhombohedral tetralayer graphene, without taking the geometric underscreening into account.}.

\begin{acknowledgments}
We thank Patrick A. Lee and Trithep Devakul for helpful discussions.
GS acknowledges support from the Walter Burke Institute for Theoretical Physics at Caltech, and from the Yad Hanadiv Foundation through the Rothschild fellowship. This work was also partially supported by the U.S.~Department of Energy, Office of Science, National Quantum Information Science Research Centers, Quantum Science Center (JA, for discussions and manuscript preparation). 

\end{acknowledgments}

\bibliographystyle{aapmrev4-2}
\bibliography{KL.bib}

\pagebreak
\begin{widetext}

\global\long\def\thesection{S.\Alph{section}}%
 \setcounter{figure}{0} 
\global\long\def\thefigure{S\arabic{figure}}%
 \setcounter{equation}{0} 
\global\long\def\theequation{S\arabic{equation}}%
\section*{Supplemental Material for ``Quantum Geometric Kohn-Luttinger Superconductivity''}

\setcounter{section}{0} \renewcommand{\thesection}{S.\arabic{section}} 
\setcounter{figure}{0} \renewcommand{\thefigure}{S\arabic{figure}} 
\setcounter{equation}{0} \renewcommand{\theequation}{S\arabic{equation}}

\setcounter{section}{0} \renewcommand{\thesection}{S.\arabic{section}} 
\setcounter{figure}{0} \renewcommand{\thefigure}{S\arabic{figure}} 
\setcounter{equation}{0} \renewcommand{\theequation}{S\arabic{equation}}

\section{Band projected Hamiltonian }
For the sake of completeness, we demonstrate the procedure of obtaining the interacting projected Hamiltonian in Eq.~(1) in the main text.
We begin with an interacting Hamiltonian with multiple bands,
\begin{equation}
    H = H_0 + H_{\rm int}.
\end{equation}
The non-interacting part of the Hamiltonian can be written as
\begin{equation}
    H_0 = \sum_{\bf k} \Psi_{{\bf k}}^\dagger {\cal H}_{{\bf k}} \Psi_{{\bf k}} = \sum_{\bf k} \psi_{a,{\bf k}}^\dagger {\cal H}_{ab,{\bf k}} \psi_{b,{\bf k}},
\end{equation}
where $\Psi_{{\bf k}}$ is a spinor of fermionic annihilation operators $\psi_{a,{\bf k}}$, which annihilate a fermion in orbital $a$ at momentum $\bf k$.
We use the convention of summing over repeated indices throughout.
The interaction Hamiltonian is
\begin{equation}
    H_{\rm int} =\frac{1}{2\Omega}\sum_{\bf q} V_{\bf q}\rho_{\bf q}\rho_{\bf -q}
\end{equation}
with $\rho_{\bf q}=\sum_{\bf k}\Psi_{{\bf k+q}}^\dagger\Psi_{{\bf k}}$, the repulsive interaction $V_{\bf q}$, and $\Omega$ is the system volume.

Let us diagonalize ${\cal H}_{\bf k}$, isolate the relevant band, and project out the rest of the degrees of freedom.
Start by diagonalizing the band
\begin{equation}
    {\cal H}_{\bf k} = {\cal U}_{\bf k}^\dagger {\cal D}_{\bf k}{\cal U}_{\bf k},
\end{equation}
where $ {\cal U}_{\bf k}$ is a unitary matrix, and the matrix ${\cal D}_{ab,{\bf k}}=\delta_{ab}\epsilon_{a,{\bf k}}$ is diagonal.
The Hamiltonian is diagonal in terms of the spinors $\Phi_{\bf k}={\cal U}_{{\bf k}}\Psi_{{\bf k}}$.
In the following, we restrict ourselves to a specific band $a=\eta$, and write the projected Hamiltonian in terms of the fermionic operators $\psi_{\bf k}\equiv \Phi_{\eta,{\bf k}}$,
\begin{equation}
    H_{\eta} = \sum_{\bf k} \left(\epsilon_{\bf k}-\mu\right)\psi_{\bf k}^\dagger\psi_{\bf k}
    +
    \frac{1}{2\Omega}\sum_{\bf q} V_{\bf q}\Tilde{\rho}_{\bf q}\Tilde{\rho}_{\bf -q}
\end{equation}
where the projected density operator is
$\tilde{\rho}_{{\bf q}}=\sum_{{\bf k}}\left[{\cal U}_{{\bf k+q}}{\cal U}_{{\bf k}}^{\dagger}\right]_{\eta\eta}\psi_{{\bf k+q}}^{\dagger}\psi_{{\bf k}}$.
Repeating the procedures for the various different electronic flavors in our model, one recovers the Hamiltonian in Eq.~(1).

\section{RPA screened interaction}
In the random phase approximation, the interaction Hamiltonian is renormalized by processes of particle-hole pair creation in the band.
For a given scattered momentum $\bf q$, a particle with momentum $\bf k+q$ and a hole with momentum $\bf k$ are created, and then annihilated.
The RPA polarization bubble thus picks up a factor $\Lambda_{{\bf k+q},{\bf k}}$ (creation) and then a factor $\Lambda_{{\bf k},{\bf k+q}}$ (annihilation), and overall the factor $\left|\Lambda_{{\bf k+q},{\bf k}}\right|^2$, see Fig.~\ref{fig:RPAdemo}.

\begin{figure}
    \centering
    \includegraphics[width=14.5cm]{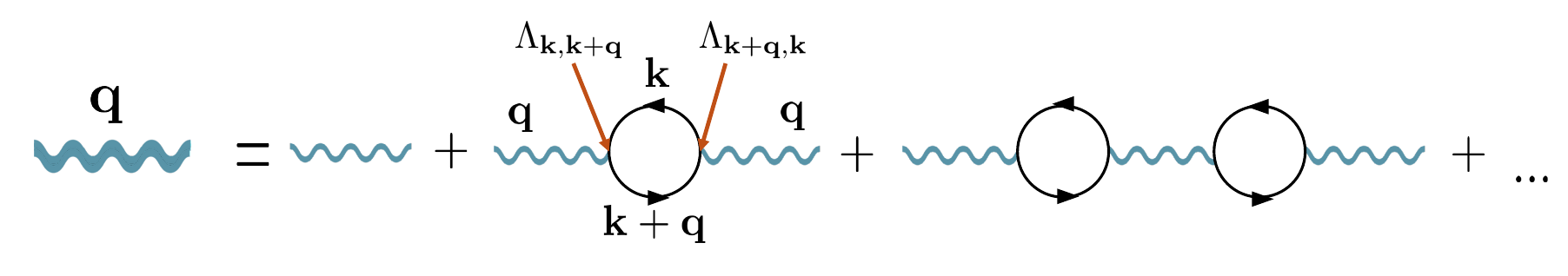}
    \caption{
    Diagrammatic representation of the RPA screened interaction.
    The thick line corresponds to the screened interaction, and thin lines are the bare interaction.
    Each internal particle-hole ``bubble'' acquires a coefficient
    $\Lambda_{{\bf k+q},{\bf k}}\Lambda_{{\bf k},{\bf k+q}}=\left|\Lambda_{{\bf k+q},{\bf k}}\right|^2$.
    }
    \label{fig:RPAdemo}
\end{figure}

The static polarization within is thus given by
\begin{equation}
    \Pi_{\bf q}= 
    - N_f \sum_{\bf k}
    \left|\Lambda_{\bf k,k+q}\right|^2
    \frac{ n\left(\xi_{\bf k+q}\right) - n\left(\xi_{\bf k}\right) }
    {\xi_{\bf k+q} - \xi_{\bf k}},
\end{equation}
where $N_f$ is the number of flavors, $\xi_{\bf k}=\epsilon_{\bf k}-\mu$, and $n\left(x\right)=1/\left(1+e^{ x/T}\right)$ is the Fermi-Dirac distribution with temperature $T$.

The RPA-screened Coulomb repulsion may be written as,
\begin{equation}
    V_{\bf q}^{\rm RPA}=\frac{V_{\bf q}}{1+\Pi_{\bf q}  V_{\bf q}}
\end{equation}
We now examine how does the screened interaction depend on momentum $\bf q$.
We examine the second derivative of the RPA-screened interaction,
\begin{align}
    \partial_{q}^{2}V_{{\bf q}}^{{\rm RPA}}	&=\partial_{q}\left(\partial_{q}V_{{\bf q}}^{{\rm RPA}}\right)
	\nonumber\\&=\partial_{q}\left(\frac{V_{{\bf q}}'-V_{{\bf q}}^{2}\Pi_{{\bf q}}'}{\left(1+\Pi_{{\bf q}}V_{{\bf q}}\right)^{2}}\right)
	\nonumber\\&=\frac{\left(V_{{\bf q}}''-V_{{\bf q}}^{2}\Pi_{{\bf q}}''\right)}{\left(1+\Pi_{{\bf q}}V_{{\bf q}}\right)^{2}}-2\frac{\left(V_{{\bf q}}'-V_{{\bf q}}^{2}\Pi_{{\bf q}}'\right)\left(\Pi_{{\bf q}}'V_{{\bf q}}+\Pi_{{\bf q}}V_{{\bf q}}'\right)+V_{{\bf q}}V_{{\bf q}}'\Pi_{{\bf q}}'\left(1+\Pi_{{\bf q}}V_{{\bf q}}\right)}{\left(1+\Pi_{{\bf q}}V_{{\bf q}}\right)^{3}}
	\nonumber\\&=\frac{\left(V_{{\bf q}}''-V_{{\bf q}}^{2}\Pi_{{\bf q}}''\right)}{\left(1+\Pi_{{\bf q}}V_{{\bf q}}\right)^{2}}-2\frac{V_{{\bf q}}'\left[\Pi_{{\bf q}}V_{{\bf q}}'+2V_{{\bf q}}\Pi_{{\bf q}}'\right]-V_{{\bf q}}^{3}\left(\Pi_{{\bf q}}'\right)^{2}}{\left(1+\Pi_{{\bf q}}V_{{\bf q}}\right)^{3}}
\end{align}
At ${\bf q} \to 0$, we can take advantage of the fact that the first derivatives of $\Pi_{\bf q}$ and $V_{\bf q}$ vanish.
Thus,
\begin{equation}
    \partial_{q}^{2}V_{{\bf q}}^{{\rm RPA}}|_{q\to0}=\frac{V_{{\bf q}}''|_{q\to0}-V_{{\bf 0}}^{2}\Pi_{{\bf q}}''|_{q\to0}}{\left(1+\Pi_{{0}}V_{{0}}\right)^{2}},
\end{equation}
which is the result we utilize in the main text.
Here, $\Pi_0$ and $V_0$ are the ${\bf q}\to 0$ limits of $\Pi_{\bf q}$ and $V_{\bf q}$, respectively.

For a rotationally invariant model with a parabolic spectrum, we may write the static polarizability in the $T=0$ limit as
\begin{align}
    \Pi^{\rm symm.}\left(q\right)
    &=
    -2\Pi_0
    \int
     k dk
    \int_0^\pi \frac{d\beta}{\pi}
    {\cal F}\left(q,k,\beta\right)
    \frac{ \Theta\left(k_F-\sqrt{k^{2}+q^2+2qk\cos\beta}\right) - \Theta\left(k_F-{k}\right) }
    {q^2+2qk\cos\beta},\label{eq:symmetricPi}
\end{align}
and we used $\left|{\bf k}+{\bf q}\right|=\sqrt{k^{2}+q^2+2qk\cos\beta}$.
In the second line, we explicitly plugged in parabolic dispersion with effective mass $m$.
Notice that the upper bound on the $k$ integration need not exceed $k<2k_F\sqrt{1+\left(q/2k_F\right)^2}+q/2$ in this case.
In our calculations for this work, we utilize Eq.~\eqref{eq:symmetricPi}, which is more numerically stable when dealing with circular Fermi surfaces.

\section{BCS gap equation}
We now plug in the RPA-renormalized Coulomb repulsion $V_{\bf q}^{\rm RPA}$ into the band-projected Hamiltonian, and seek a self-consistent solution to the BCS gap equation,
\begin{equation}
    \Delta_{\bf k}=
    -\int\frac{d{\bf k'}}{\left(2\pi\right)^2}
    u_{\bf kk'}
    \frac{\tanh{\frac{E_{\bf k'}}{2T}}}{2E_{\bf k'}}
    \Delta_{\bf k'}.
\end{equation}
The conventions here is that positive $u$ corresponds to repulsive interaction.
Notice, for constant $\Delta_{\bf k}$ and constant attraction (negative $u$), one recovers
\begin{equation}
    \frac{1}{\left|u\right|}=
    \int\frac{d{\bf k'}}{\left(2\pi\right)^2}
    \frac{\tanh{\frac{E_{\bf k'}}{2T}}}{2E_{\bf k'}}.\label{eq:BCS}
\end{equation}

In our scenario, the Copper channel interaction is comprised of two parts,
\begin{equation}
    u_{\bf kk'} =\Lambda_{{\bf kk'}}^{\left(1\right)}\Lambda_{{\bf kk'}}^{\left(2\right)} V_{\bf k-k'}^{\rm RPA},
\end{equation}
where we have provided an upper index to the form factors in order to distinguish contributions from the two electrons forming the Cooper pair.
The physical meaning of these factors appearing in the interaction is clear.
The Cooper interaction scatters a pair of particles into another pair.
The form factors simply indicate the overlap between the initial and final states.

In case of time-reversal symmetric superconductor, e.g., if electrons come from opposite time-reversed valleys, $\Lambda_{{\bf kk'}}^{\left(2\right)}=\left[\Lambda_{{\bf kk'}}^{\left(1\right)}\right]^*$, and one simply has
$u_{\bf kk'}={\cal F}_{{\bf k},{\bf k'}}V_{\bf k-k'}^{\rm RPA}$.
However, if superconductivity pairs electron from the same time-reversal symmetry breaking band, 
$u_{\bf kk'}=\Lambda_{{\bf kk'}}^2 V_{\bf k-k'}^{\rm RPA}$.

We can simplify a bit, by approximating the radial dependence of $u_{\bf kk'}$ as being roughly constant.
This is justified, since the main contributions in Eq.~\eqref{eq:BCS} come from the vicinity of the Fermi surface.
Taking also the linearized version, $E_{\bf k'}=\xi_{\bf k}=\frac{k'^{2}-k_{F}^{2}}{2m}$, we find,
\begin{equation}
    \Delta_{\theta}=-I\left(T_c\right)\int\frac{d\theta'}{2\pi}u_{{\bf \theta\theta'}}\Delta_{{\bf \theta'}}.
\end{equation}
with
\begin{equation}
    I\left(T_c\right) = -\frac{1}{4\pi}\int_0^{\Lambda} kdk\frac{\tanh\frac{\frac{k^{2}-k_{F}^{2}}{2m}}{2T_{c}}}{\frac{k^{2}-k_{F}^{2}}{2m}},
\end{equation}
Where $\Lambda$ is a momentum cutoff for the integration.
After a change of variables, the integral becomes ($E_\Lambda=\Lambda^2/\left(2m\right)-E_F$),
\begin{equation}
    I\left(T_c\right) =-\frac{m}{4\pi}\int_{-E_{F}/2T_{c}}^{E_\Lambda/2T_{c}}dx\frac{\tanh x}{x}.
\end{equation}
Anticipating $T_c$ to be much smaller compared to the band energy scales, we approximate the integral in the standard fashion,
\begin{equation}
    I\left(T_c\right) \approx -\frac{m}{4\pi}\left[\log\left(\frac{E_\Lambda}{2T_{c}}\right)+\log\left(\frac{E_{F}}{2T_{c}}\right)+C\right],
\end{equation}
where $C$ is some $O\left(1\right)$ constant.
Collecting the contributions into one effective energy scale, $W\equiv\sqrt{E_\Lambda E_{F}e^{C}}/2$,

\begin{align}
    \Delta_{\theta}&=-\frac{m}{2\pi}\log\frac{W}{T_c}\int\frac{d\theta'}{2\pi}u_{\left[\theta,\theta'\right]_{k_F}}\Delta_{\theta'}\nonumber\\
    &=-\log\frac{W}{T}{\cal K}_{\left[\theta,\theta'\right]_{k_F}}\Delta_{\theta'},
\end{align}
with the linear operator
\begin{equation}
    {\cal K}_{\left[\theta,\theta'\right]_{k_F}}\Delta_{\theta'}=\frac{m}{2\pi}\int\frac{d\theta'}{2\pi}u_{\left[\theta,\theta'\right]_{k_F}}\Delta_{\theta'}.\label{eq:Klinearoperatorbcs}
\end{equation}
Here, we use the convention 
$u_{\left[\theta,\theta'\right]_{k_F}}$ to represent the interaction between to points on the Fermi surface (with momentum $k_F$) at different radial angles, $\theta$ and $\theta'$.
The leading superconducting instability corresponds to \textit{the most negative eigenvalue} of $\cal K$, which we denote as $\lambda$.
Then, the critical temperature is
$T_c\approx W \exp\left(-\frac{1}{\lambda}\right)$.

\subsection{Time-reversal broken superconductor}
As discussed above, in cases where the parent normal state breaks time-reversal symmetry, the BCS gap equation may be modified in significant ways.
For concreteness, let us assume rotational invariance as we have for the models considered in the main text.
Assuming the same decomposition for the RPA interaction itself,
$V^{k_F}=v_{0}/2+\sum_{n>0}v_{n}\cos\left(n\theta\right)$, we can similarly decompose the form-factor contribution,
\begin{equation}
    F^{k_F}=\frac{f_0}{2}+\sum_m
    \left(
    a_m e^{im\theta} + b_m e^{-im\theta}
    \right),
\end{equation}
where time-reversal breaking implies that generally $a_m\neq b_m$.
The effective interaction thus decomposes to
\begin{align}
    F^{k_{F}}V^{k_{F}}&=
    \frac{f_{0}v_{0}}{4}+\sum_{m>0}\frac{f_{0}v_{m}}{4}\left[e^{im\theta}+e^{-im\theta}\right]\nonumber\\
    &+\sum_{m>0,n\geq0}\left[\frac{a_{m}v_{n}}{2}\left(e^{i\left(m+n\right)\theta}+e^{i\left(m-n\right)\theta}\right)+\frac{b_{m}v_{n}}{2}\left(e^{-i\left(m+n\right)\theta}+e^{-i\left(m-n\right)\theta}\right)\right].
\end{align}
Each term $e^{i\ell\theta}$, with $\ell$ an integer which may be positive or negative, corresponds to an eigenvector of the operator in Eq.~\eqref{eq:Klinearoperatorbcs} with a gap function proportional to $e^{i\ell\theta}$.

Let us consider a concrete example, from the topological band inversion model (see Sec.~\ref{app:TBI}).
Assume that the quantum-metric induced overscreening is strong enough such that the screened interaction develops a strong p-wave instability,
$V^{k_F}_{{\rm BHZ}}=v_{0}/2-\left|v_{1}\right|\cos\theta$.
Clearly, $v_{0}/2>\left|v_{1}\right|$, since the interaction does not change sign within the RPA treatment.
If pairing occurs between bands of the same time-reversal flavors, one obtains
\begin{equation}
    F^{k_F}_{{\rm BHZ}}=\frac{1+2\kappa^{2}e^{i\theta}+\kappa^{4}e^{2i\theta}}{\left(1+\kappa^{2}\right)^{2}}.
\end{equation}
The effective interaction is then decomposed as ($f_0=\frac{2}{\left(1+\kappa^{2}\right)^{2}}$, $a_1=\frac{2\kappa^{2}}{\left(1+\kappa^{2}\right)^{2}}$, $a_2=\frac{\kappa^{4}}{\left(1+\kappa^{2}\right)^{2}}$),
\begin{align}
F^{k_{F}}_{{\rm BHZ}}V^{k_{F}}_{{\rm BHZ}} & =\frac{f_{0}v_{0}}{4}-\frac{a_{1}\left|v_{1}\right|}{2}\nonumber\\
 & +\left(\frac{a_{1}v_{0}}{2}+\frac{a_{2}v_{1}}{2}+\frac{f_{0}v_{1}}{4}\right)e^{i\theta}+\frac{f_{0}v_{1}}{4}e^{-i\theta}\nonumber\\
 & +\left(\frac{a_{2}v_{0}}{2}+\frac{a_{1}v_{1}}{2}\right)e^{2i\theta}\nonumber\\
 & +\frac{a_{2}v_{1}}{2}e^{3i\theta}.
\end{align}
In the vicinity of $\kappa\approx 1$ the two leading instabilities are proportional to $e^{-i\theta}$ (``$p_{-}$'', leading for $\kappa<1$), or to $e^{3i\theta}$ (``$f_{+}$'', leading for $\kappa>1$).
It is thus clear that relaxation of the time-reversal symmetry opens the door to an even richer range of possibilities, and generically produce chiral superconductivity as the leading gap order parameter.

\subsection{Isotropic and homogeneous quantum metric}
Let us briefly discuss a specific non-generic form of the quantum metric and its affect on the effective coupling interaction $u_{\bf k,k'}$.
In the case where the quantum metric is isotropic and homogeneous, as in, e.g., a Landau level of electrons in a strong magnetic field, the norm of the form factors depends solely on the distance in the Brillouin zone.
Namely,
\begin{equation}
    \left|\Lambda_{\bf k,k'}\right|^2 = e^{-\frac{1}{2}{\cal B}\left|{\bf k-k'}\right|^2},
\end{equation}
where ${\cal B}>0$ is the absolute value of the uniform Berry curvature.
Calculating the static polarization then becomes trivial, 
$\Pi_{\bf q}=e^{-\frac{1}{2}{\cal B}\left|{\bf q}\right|^2}\Pi^{\rm trivial}$, where $\Pi^{\rm trivial}$ is the polarization function for the same band with trivial quantum geometry.
The BCS interaction $u_{\bf k k'}$ also simplifies to  $u_{\left|{\bf k k'}\right|}\equiv u_{\left|{\bf q}\right|}=e^{-\frac{1}{2}{\cal B}q^{2}}\frac{V_{q}}{1+e^{-\frac{1}{2}{\cal B}q^{2}}\Pi^{{\rm trivial}}V_{q}}$.
One may directly probe its properties at $\bf q\to 0$ (assuming for simplicity that $\Pi^{\rm trivial}$ is constant at low momenta, which is accurate for 2D electrons with parabolic dispersion),
\begin{equation}
    \partial_{q}^{2}u_{{\bf q}}|_{q \to 0}=
    -V_{0}\frac{\frac{2}{3}d^{2}+{\cal B}}{\left(1+\Pi^{{\rm trivial}}V_{0}\right)^{2}}<0.
\end{equation}
Clearly, in this case the quantum geometric contribution from $\cal B$ only acts to make the interaction drop even more rapidly with increasing momentum.
We thus conclude that a metric which is both isotropic and homogeneous, i.e., the quantum distance solely depends on the relative momentum distance $\left|{\bf k-k'}\right|$, does not lead to pronounced enhancement of Kohn-Luttinger superconductivity.

\section{Tunable metric model}
Let us introduce the model introduced in Ref.~\cite{tunableMetricPRL_PhysRevLett.130.226001}, and its extension which we utilize to model a metric-tunable Fermi liquid.
The model has two orbital degrees of freedom, addressed by Pauli matrices $\tau_i$, and its two flavors are addressed by $\sigma_i$.
The momentum space Hamiltonian is
\begin{equation}
    H_{\rm tun.}=\frac{E_g}{2}\sum_{{\bf k}}c_{{\bf k}}^{\dagger}\left[\sin\left(\zeta{\ensuremath{\alpha_{{\bf k}}}}\right)\tau_{x}+\sigma_{z}\cos\left(\zeta{\ensuremath{\alpha_{{\bf k}}}}\right)\tau_{y}\right]c_{{\bf k}}
    -t\sum_{{\bf k}}c_{{\bf k}}^{\dagger}
    \left[\tau_0\sigma_0\alpha_{\bf k}\right]
    c_{{\bf k}},\label{eq:Htunablemetric}
\end{equation}
where $c_{\bf k}$ is a spinor of fermionic annihilation operator at momentum $\bf k$, and ${\alpha_{{\bf k}}}=\left[\cos\left(k_{x}a\right)+\cos\left(k_{y}a\right)\right]$, and $a$ is the square lattice constant.
Notice that the first part of this Hamiltonian does not disperse at all as a function of momentum, regardless of the parameter $\zeta$. 
It only serves as a ``blueprint'' for the resulting bands quantum geometry.
The part of $H_{\rm tun.}$ which is proportional to $t$ represents nearest-neighbor hopping on the square lattice, without a change to the orbital or flavor indices.
This term does not contribute or affect the quantum geometrical properties.
We consider the limit $t\ll E_g$ where the valence and conduction bands are well-separated by a constant gap of $E_g$.
When the Fermi level is close to either the top or the bottom of one of the bands, we may approximate the spectrum as parabolic, with effective mass $m_{\rm eff.}\approx\left(a^2 t\right)^{-1}$.
Without loss of generality, we assume the Fermi surface lies near the bottom of the valence band.

The eigenvector at momentum $\bf k$, at band $\pm$ is given by
\begin{equation}
    u_{{\bf k}}=\frac{1}{\sqrt{2}}e^{i\sigma\alpha_{{\bf k}}\frac{\tau_{z}}{2}}\left(1,\pm i\right)=\frac{1}{\sqrt{2}}\begin{pmatrix}e^{i\sigma\frac{\alpha_{{\bf k}}}{2}}\\
\pm ie^{-i\sigma\frac{\alpha_{{\bf k}}}{2}}
\end{pmatrix}.
\end{equation}
The form factors are thus (setting $a=1$ for convenience),
\begin{equation}
    \Lambda_{\bf k+q,k}=
    \left\langle 
    u_{{\bf k+q}}|u_{{\bf k}}\right\rangle 
    =
    \cos\zeta\frac{\alpha_{{\bf k}}-\alpha_{{\bf k+q}}}{2}
    =\cos\zeta\left[
    \sin\frac{q_{x}}{2}\sin\left(k_{x}+\frac{q_{x}}{2}\right)+\sin\frac{q_{y}}{2}\sin\left(k_{y}+\frac{q_{y}}{2}\right)
    \right].
\end{equation}
In the approximation where the Fermi surface lies near the bottom or the top of the band, one recovers rotational symmetry in the Brillouin zone, allowing us to approximate,
\begin{equation}
    \left| \Lambda_{\bf k+q,k}\right|^2
    \approx
    {\cal F}\left(q,k,\beta\right)
    =
    \cos^{2}\left[\frac{\zeta}{4}\left(2kq\cos\beta+q^{2}\right)\right],\label{eq:tunablemetricrotationalff}
\end{equation}
where we denoted ${\bf k}=k\left(\cos \beta,\sin\beta\right)$ and ${\bf q}=q\left(1,0\right)$.

In the general case, without restricting to the top/bottom of the band, one may expand the form factors expression for small momenta transfer, in order to extract the quantum metric $g_{\mu\nu}$,
\begin{align}
    \left| \Lambda_{\bf k+q,k}\right|^2
    &=\cos^2\left[\zeta\left[
    \sin\frac{q_{x}}{2}\sin\left(k_{x}+\frac{q_{x}}{2}\right)+\sin\frac{q_{y}}{2}\sin\left(k_{y}+\frac{q_{y}}{2}\right)\right]\right]\nonumber\\
    &\approx 1 - \frac{\zeta^2}{4} \left[
    q_{x}\sin k_{x}+q_{y}\sin k_{y}\right]^2\nonumber\\
    &= 1- \frac{\zeta^2}{4}
    \left(q_x^2\sin^2 k_x +q_y^2\sin^2 k_y\right)
    -\frac{\zeta^2}{2}q_x q_y \sin k_x \sin k_y.
\end{align}
Restoring $a$ for correct units, we find
\begin{equation}
    g_{\mu\nu}\left({\bf k}\right) = \frac{\left(\zeta a\right)^2}{4} \sin\left(k_\mu a\right)\sin\left(k_\nu a\right).\label{Eq:tunablemetricresult}
\end{equation}
This allows us to extract $\bar{g}$ by integrating over the circular Fermi surface,
\begin{equation}
    \bar{g}=\frac{\left(\zeta a\right)^2}{4}
    \int \frac{d\theta}{2\pi}
    \sin^2 \left(k_F a \cos\theta\right)
    =\frac{\left(\zeta a\right)^2}{8}
    \left[1-J_0\left(2k_Fa\right)\right]
    \to \frac{\left(\zeta a\right)^2}{8} \left(k_F a\right)^2,
\end{equation}
where in the last step we used the assumption of Fermi energy in the vicinity of the bottom of one of the bands.

It is useful to convert the metric obtained in Eq.~\eqref{Eq:tunablemetricresult} in Cartesian coordinates to cylindrical ones, focusing on areas near the bend edges.
There, 
$g_{xx}=\frac{\left(\zeta a\right)^2}{4}x^{2}$, $g_{yy}=\frac{\left(\zeta a\right)^2}{4}y^{2}$, $g_{xy}=\frac{\left(\zeta a\right)^2}{4}xy$, one finds the simplified form of the line element
\begin{equation}
     ds^2 \approx r^{2}\frac{\left(\zeta a\right)^2}{4}dr^{2},
\end{equation}
where $r$ is the distance from the origin (or, if alternatively we focus on an almost full band, distance from ${\bf k}a=\left(\pi,\pi\right)$).
Notice that the line-element defined by this metric is independent of displacement along an equal-$r$ contour, i.e., the quantum-metric-distance between points on the Fermi surface is identically zero.
We emphasize that this is not an artifact of the small-momentum approximation we employ, but rather a direct consequence of the structure in Eq.~\eqref{eq:Htunablemetric}. 
Namely, the band form factors and the Fermi surface have a similar functional dependence on $\alpha_{\bf k}$.
As a result, states with the equal energy are described by identical wavefunctions $u_{\bf k}$.
We note that the symmetric form of Eq.~\eqref{eq:tunablemetricrotationalff} maintains this quality as well.




\section{Topological band inversion model}\label{app:TBI}
Let us consider the following staple in the field of topological insulators: the low-energy theory of topological band inversion~\cite{QSHbernevig}.
The Hamiltonian is
\begin{equation}
    H_{\rm BHZ}=\sum_{{\bf k}}\Psi_{{\bf k}}^{\dagger}\left[\
    vk_x\sigma_x
    +vk_y\sigma_y
    +\left(\frac{k^2}{2m}+M\right)\sigma_z
    \right]\Psi_{{\bf k}},
\end{equation}
with $\Psi_{\bf k}$ a spinor of fermionic annihilation operators at momentum $\bf k$ living in the Hilbert space spanned by the Pauli matrices $\sigma_i$.
We make the judicious choice of $M=-mv^2/2$, leading to an exactly parabolic spectrum of the two bands,
\begin{equation}
    \epsilon_{\bf k}=\pm\frac{k^2+Q^2}{2m},
\end{equation}
where we defined $Q=mv$.
The wavefunction of one of the bands is
\begin{equation}
    |u_{{\bf k}}\rangle=\frac{1}{\sqrt{k^{2}+Q^{2}}}\begin{pmatrix}Q\\
k_{x}+ik_{y}
\end{pmatrix},
\end{equation}
such that
\begin{equation}
    \left\langle u_{{\bf k'}}|u_{{\bf k}}\right\rangle 
    =\frac{Q^{2}+{\bf k}\cdot{\bf k'}+i{\bf k}\times{\bf k'}}{\sqrt{k^{2}+Q^{2}}\sqrt{k'^{2}+Q^{2}}}.
\end{equation}
Exploiting the rotational symmetry inherent to this model, we may compute the form factors similarly to Eq.~\eqref{eq:tunablemetricrotationalff},
\begin{equation}
    {\cal F}\left(q,k,\theta\right)=\left|\frac{Q^{2}+k^{2}+qke^{-i\theta}}{\sqrt{k^{2}+Q^{2}}\sqrt{k^{2}+q^{2}+2kq\cos\theta+Q^{2}}}\right|^{2}=\frac{\left(k^{2}+Q^{2}\right)^{2}+q^{2}k^{2}+2\left(k^{2}+Q^{2}\right)qk\cos\theta}{\left(k^{2}+Q^{2}\right)\left(k^{2}+Q^{2}+q^{2}+2kq\cos\theta\right)}.
\end{equation}
We note that for small $q$, one finds
\begin{equation}
    {\cal F}\left(q,k,\theta\right)\approx1-q^{2}\frac{Q^{2}}{\left(k^{2}+Q^{2}\right)^{2}}+{\cal O}\left(q^{3}\right)
    \approx 1-\left(\frac{q}{2k_F}\right)^2
    \frac{4\kappa^2}{\left(\kappa^2+1\right)^2},
\end{equation}
with the definition $\kappa\equiv k_F/Q$.
We thus recover the isotropic (yet inhomogenous) metric $g_{\rm BHZ}\equiv\frac{Q^{2}}{\left(k^{2}+Q^{2}\right)^{2}}$, which is readily averaged over the Fermi surface,
\begin{equation}
    \bar{g}=\frac{Q^{-2}}{\left(\kappa^{2}+1\right)^{2}}.
\end{equation}

Let us compute the form factors on the Fermi surface itself (which pay a role in the BCS gap equation),
\begin{align}
{\cal F}_{k_{F}}\left(\theta\right) & =\left|\left\langle u_{k_{F}\left(\cos\theta,\sin\theta\right)}|u_{k_{F}\left(1,0\right)}\right\rangle \right|^{2}\nonumber\\
 & =\left|\frac{Q^{2}+k_{F}^{2}e^{i\theta}}{k_{F}^{2}+Q^{2}}\right|^{2}\nonumber\\
 & =\frac{Q^{4}+k_{F}^{4}+2Q^{2}k_{F}^{2}\cos\theta}{\left(k_{F}^{2}+Q^{2}\right)^{2}}\nonumber\\
 & =\frac{\kappa^{4}+1+2\kappa^{2}\cos\theta}{\left(\kappa^{2}+1\right)^{2}},
\end{align}
where ${\cal F}_{k_{F}}\left(\theta\right)$ represents the overlap between two points with Fermi momentum $k_F$ at a relative angle $\theta$.

\section{Quantum geometric enhancement of conventional superconductivity}
We have demonstrated throughout this work that due to unique screening that occurs in the presence of non-trivial quantum geometry and Fubini-Study metric, an effective attraction may develop in one of the angular channels of the on-Fermi-surface interaction.
However, the conventional s-wave channel, corresponding to the coefficient $u_0$ in the angular decomposition, is significantly modulated by quantum geometric effects as well.

For concreteness, consider the results we have obtained for the BHZ model.
The expression for $u_0$ takes the form of
\begin{equation}
    u_{0}=\left(1-2\bar{f}\right)v_{0}-2\bar{f}\left|v_1\right|.
\end{equation}
As a reminder, $\bar{f}=\frac{\kappa^2}{\left(\kappa^2+1\right)^2}$ originate in the Fermi surface form factors, and $v_0$, $v_1$ are the first angular harmonics of the screened RPA interaction.
Generically in our results we find that $v_1<0$, as implied in the expression for $u_0$.
In Fig.~\ref{fig:andersonmoreltcfig}a we plot $u_0$ as a function of $\kappa$ (the normalized Fermi momentum).
Notably, due to the negative contribution by the p-wave component $v_1$, the coefficient $u_0$ is much lower around the ``optimal'' $\kappa=1$.

The interaction in this channel is clearly still positive, and as such cannot lead to a superconducting instability on its own.
Nevertheless, in a more conventional scenario where superconducting pairing is mediated by some retarded attractive interaction $g_{\rm pair}$, which becomes effective at energy scales below $\omega^*$, the quantum geometric effect can become significant.
(The obvious example is that of interactions mediated by electron-phonon interaction.)
To explore such a scenario, we employ the Tolmachev-Anderson-Morel framework~\cite{tolmachev1962logarithmic,AndersonMorel}.

In the presence of the bare (though quantum-metric modified) repulsion $u_0$, we integrate out the fermionic degrees of freedom from an initial cutoff $W$ down to $\omega^*<W$.
One recovers the effective renormalized repulsion at this scale, the so-called Anderson-Morel pseudopotential,
\begin{equation}
    \mu^*=\frac{u_0}{1+\Pi_0 u_0 \log\frac{W}{\omega^*}}.
\end{equation}
Introduction of an attractive interaction which is strong enough to overcome $\mu^*$, i.e., $\left|g_{\rm pair}\right|>\mu^*$, one recovers an estimate of the superconducting transition temperature $T_c$,
\begin{equation}
    k_B T_c\approx \omega^* \exp\left[-\frac{1}{\left(\left|g_{\rm pair}\right|-\mu^*\right)\Pi_0}\right].
\end{equation}

\begin{figure}
    \centering
    \includegraphics[width=14.5cm]{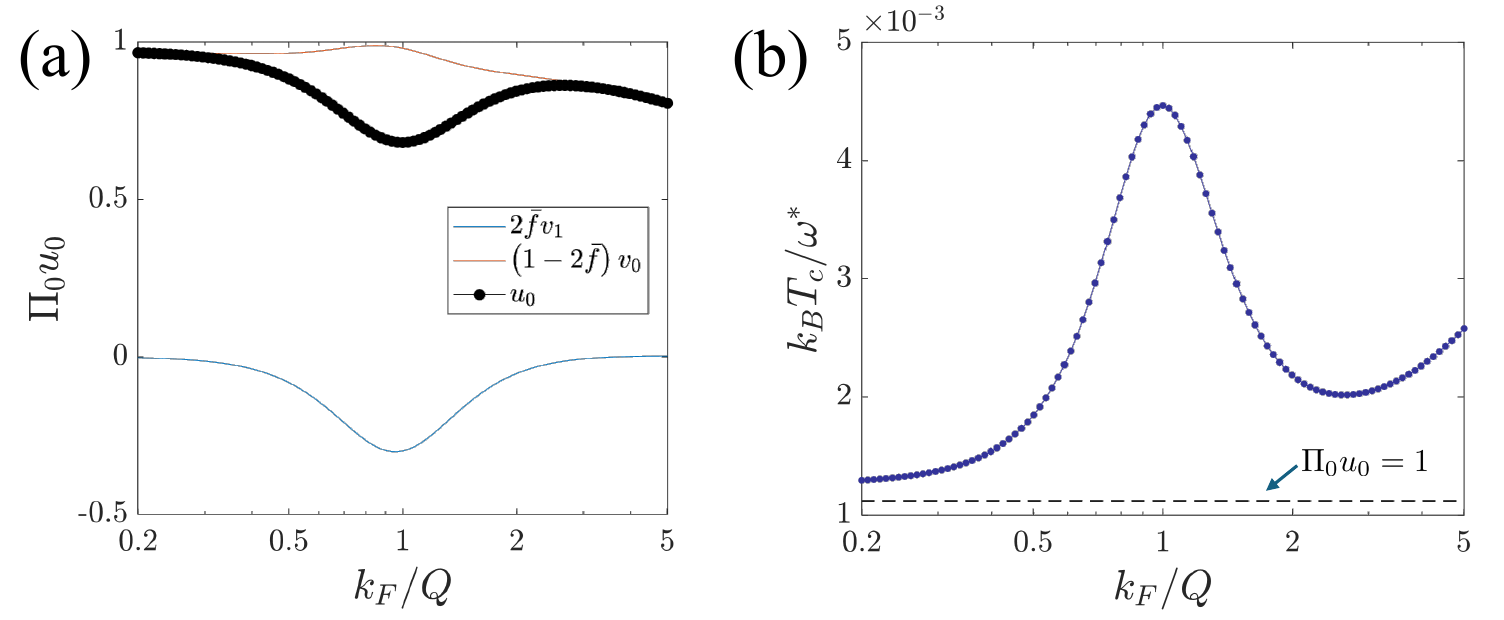}
    \caption{
    Conventional superconductivity enhanced by quantum geometric effects.
    (a)
    The s-wave coupling constant $u_0$ (black) and its decomposition to components of the screened interaction $v_0$ (red) and $v_1$ (blue).
    We use the same parameters as in Fig.~3b in the main text.
    The repulsion is weakest around the optimal value of the Fermi momentum $k_F\approx Q$.
    (b)
    The calculated $T_c$ from the Anderson-Morel mechanism of retarded attractive interactions (purple).
    The dashed line corresponds to a geometrically trivial band where $u_0$ is not modulated by the quantum metric (we choose $\Pi_0 u_0=1$ for comparison).
    We use 
    $\Pi_0 \left|g_{\rm pair}\right|=0.45$, 
    and $W=10\omega^*$.
    }
    \label{fig:andersonmoreltcfig}
\end{figure}

The effect of quantum geometry on the conventional $T_c$ in this retardation scenario are shown in Fig.~\ref{fig:andersonmoreltcfig}b.
The dip in $u_0$ (seen in Fig.~\ref{fig:andersonmoreltcfig}a) leads to a marked enhancement of $T_c$ near the optimal regime.
Notably, as a function of density the superconducting $T_c$ changes by roughly $\sim 300\%$, whereas the density of states remains perfectly constant.
The system is no more closer to a van Hove singularity or to some nesting condition, nor does it change its Fermi-surface topology -- \textit{this giant modulation is purely a quantum geometric effect}.

\section{Additional figures}
For the calculations regarding the TM model, we plot the superconductivity instability order $\tilde{n}$ appropriate for 
Fig.~2c
(also shown in Fig.~\ref{fig:extrafigs}a for convenience) in Fig.~\ref{fig:extrafigs}b.

For the BHZ model, we focus on the optimal $\kappa=1$ regime, and plot the dependence of the coupling constant on distance to the gates $d$ and on the Thomas-Fermi wavelength $\lambda_{\rm TF}$ in Fig.~\ref{fig:extrafigs}c.
As one may expect, since $\lambda_{\rm TF}\propto\epsilon$ the dielectric constant, and since superconductivity is Coulomb-repulsion mediated, superconductivity dramatically increases at small values of $\lambda_{\rm TF}$.

The dependence on $d$, however, is much less pronounced, as demonstrated in Fig.~\ref{fig:extrafigs}d.
The $d$-dependence becomes noticeable only when $d$ becomes comparable to the inter-particle separation, i.e., $k_F d\approx 1$ (notice that in Fig.~\ref{fig:extrafigs} $k_F=Q$).

\begin{figure}
    \centering
    \includegraphics[width=18cm]{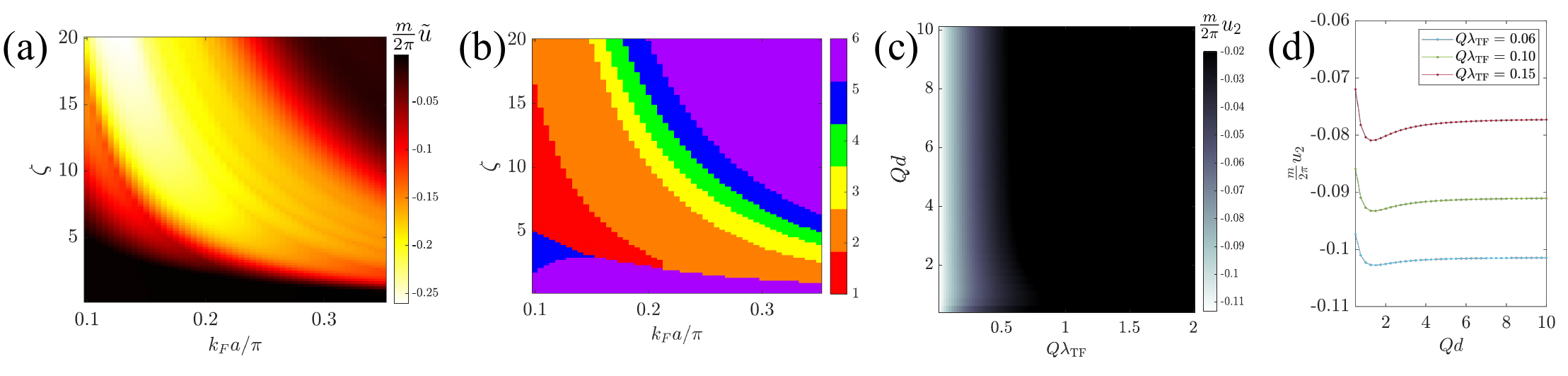}
    \caption{
    (a)
    The same data appearing in 
    Fig.~2c
    in the main text, coupling constant for the tunable metric model.
    (b)
    The corresponding $\tilde{n}$, order of the leading superconducting instability, for the data in (a).
    (c)
    Dependence of the superconducting coupling constant in the topological band inversion model on $d$ and $\lambda_{\rm TF}$, taken at $\kappa=1$.
    (d)
    Several cuts from (c), as indicated by the legend.
    }
    \label{fig:extrafigs}
\end{figure}

\section{Rhombohedral graphene}
In order to demonstrate the potential relevance of the geometric screening effect to recent superconductivity discoveries in rhombohedral graphene, we present in Fig.~\ref{fig:graphenegeo} the calculation of the Fubini-Study metric trace ${\rm tr} g$ for two representative case of Bernal stacked bilayer graphene and rhombohedral trilayer graphene.

For the band calculations for the bilayer graphene near one of the graphene valleys $K/K'$ points in momentum space, we use the Hamiltonian~\cite{mccannElectronicPropertiesBilayer2013} 
\begin{equation}
H_{BLG}=\sum_{\mathbf{k},\tau,s}c_{\tau s\mathbf{k}}^{\dagger}h_{\tau}\left(\mathbf{k}\right)c_{\tau s\mathbf{k}},\label{eq:BLGhAMILTONIAN}
\end{equation}
with $c_{\tau s\mathbf{k}}=\left(A_{1,\tau s\mathbf{k}},B_{1,\tau s\mathbf{k}},A_{2,\tau s\mathbf{k}},B_{2,\tau s\mathbf{k}}\right)^{T},$
where $X_{i,\tau s\mathbf{k}}$ annihilates an electron on sub-lattice
$X$ in layer $i$, with spin $s$, and momentum $\mathbf{k}$ near
the valley $\tau$. The matrix $h_{\tau}$ is given by
\begin{equation}
h_{\tau}\left(\mathbf{k}\right)=\begin{pmatrix}\frac{U}{2} & v_{0}\Pi_{\tau}^{*} & -v_{4}\Pi_{\tau}^{*} & -v_{3}\Pi_{\tau}\\
v_{0}\Pi_{\tau} & \frac{U}{2}+\Delta' & \gamma_{1} & -v_{4}\Pi_{\tau}^{*}\\
-v_{4}\Pi_{\tau} & \gamma_{1} & -\frac{U}{2}+\Delta' & v_{0}\Pi_{\tau}^{*}\\
-v_{3}\Pi_{\tau}^{*} & -v_{4}\Pi_{\tau} & v_{0}\Pi_{\tau} & -\frac{U}{2}
\end{pmatrix},\label{eq:H04by4basis}
\end{equation}
with $\Pi_{\tau}=\tau k_{x}+ik_{y}$, and the parameters $v_{i}=\frac{\sqrt{3}}{2}a\gamma_{i},$
$a=0.246$ nm, $\gamma_{0}=2.61$ eV, $\gamma_{1}=361$ meV, $\gamma_{3}=283$
meV, $\gamma_{4}=138$ meV, and $\Delta'=15$ meV\cite{jungAccurateTightbindingModels2014}.
The interlayer potential difference $U$ is approximately $U\approx-dD/\epsilon$,
where the interlayer distance is $d\approx0.33$nm, $\epsilon\approx4.3$,
and $D$ is the displacement field. 
We diagonalize $H_{BLG}$ at each momentum, and extract the dispersion relation of the lowest-lying valence band, which we denote by $\epsilon^{\left(2\right)}_{\tau,\mathbf{k}}$.
For this band, we compute the trace of the quantum metric, and plot it in Fig.~\ref{fig:graphenegeo}a.

For rhombohedral trilayer graphene calculations, we diagonalize the Hamiltonian~\cite{ABCparametersTB_mccan,ABCparametersTB},
\begin{equation}
    h_\tau\left({\bf k}\right)=\begin{pmatrix}\Delta+\delta_{1}+\delta_{2} & \frac{\gamma_{2}}{2} & v_{0}\pi^{\dagger} & v_{4}\pi^{\dagger} & v_{3}\pi & 0\\
\frac{\gamma_{2}}{2} & -\Delta+\delta_{1}+\delta_{2} & 0 & v_{3}\pi^{\dagger} & v_{4}\pi & v_{0}\pi\\
v_{0}\pi & 0 & \Delta+\delta_{2} & \gamma_{1} & v_{4}\pi^{\dagger} & 0\\
v_{4}\pi & v_{3}\pi & \gamma_{1} & -2\delta_{2} & v_{0}\pi^{\dagger} & v_{4}\pi^{\dagger}\\
v_{3}\pi^{\dagger} & v_{4}\pi^{\dagger} & v_{4}\pi & v_{0}\pi & -2\delta_{2} & \gamma_{1}\\
0 & v_{0}\pi^{\dagger} & 0 & v_{4}\pi & \gamma_{1} & -\Delta+\delta_{2}
\end{pmatrix},
\end{equation}
where $\Delta$ is the interlayer potential difference (proportional to the displacement field $D$.
Notice we have written the Hamiltonian in the basis $\left(A_1,B_3,B_1,A_2,B_2,A_3\right)$, where $A_j/B_j$ correspond to different graphene sublattices in layer $j$.
Here, we use $\gamma_0=3.1$ eV, $\gamma_1=0.38$ eV, $\gamma_2=-15$ meV, $\gamma_3=-0.29$ eV, $\gamma_4=-141$ meV, $\delta_1=-10.5$ meV, $\delta_2=-2.3$ meV.
Here, we are interested in the dispersion of the lowest lying valence band $\epsilon^{\left(3\right)}_{\tau,\mathbf{k}}$, and its quantum metric, plotted in Fig.~\ref{fig:graphenegeo}b.

\begin{figure}
    \centering
    \includegraphics[width=18cm]{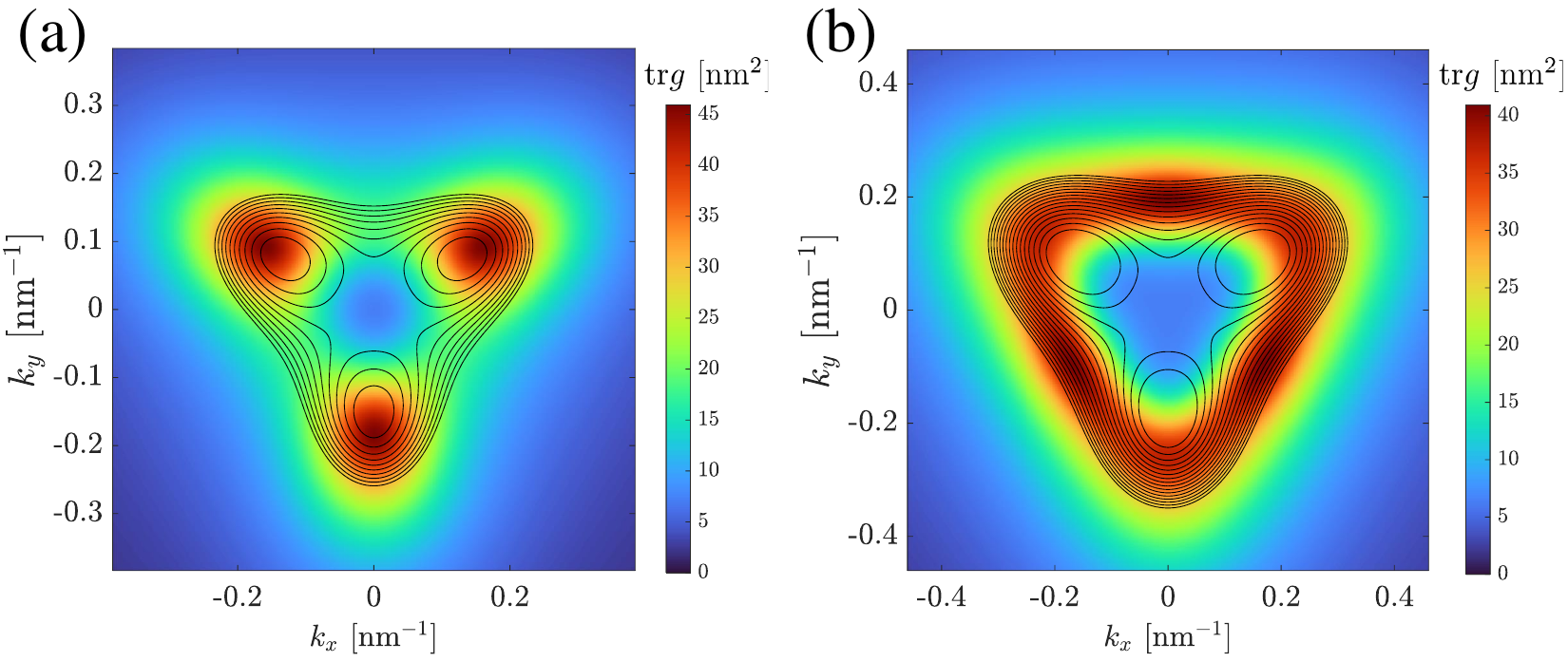}
    \caption{
    (a)
    Trace of the Fubini-Study quantum metric around on of the graphene valleys in momentum space for the lowest lying valence band of Bernal bilayer graphene, with interlayer potential difference $U=60$ meV.
    The black lines are equal energy contours (Fermi surfaces) in the range $[-36, -26]$ meV, in $1$ meV steps.
    (b)
    Similar plot for rhombohedral trilayer graphene, with $\Delta=25$ meV.
    The Fermi level contours are in the range $[-43, -28]$ meV in $1$ meV steps.
    }
    \label{fig:graphenegeo}
\end{figure}

It is quite clear that the strength of the quantum metric around the relevant Fermi levels is rather large, with a conservative order of $30$ nm$^2$.
Conservatively estimating the density of states per flavor, $m/\left(2\pi\right)\to2\cdot10^{10}$ mev$^{-1}$cm$^{-2}$, $\epsilon=12\epsilon_0$, we get
$\lambda_{\rm TF}\approx0.5$nm.
As $d$ is typically of the order of $10$ nm,
the heuristic criterion $\bar{g}>\frac{\lambda_{TF}d}{3N_f}$ is comfortably satisfied.
Furthermore, we can get a rough estimate of the quantum geometric parameter $Q$ by noticing that for a gapped Dirac cone (which are the building blocks of the gate-biased bilayer graphene band structure) $Q^{-2}$ is the maximal value of the metric.
Thus, one estimates the appropriate $Q\lambda_{\rm TF}\lesssim 0.1$, as we employed in our calculations for the main text.

Given characteristic experimental $T_c$s of the order $\sim 100$ mK, and characteristic Fermi energies of $\sim 10$ meV, one concludes that an appropriate dimensionless coupling constant (in the language used in the main text, ($\frac{m}{2\pi}\tilde{u}$) should be of order $0.1$--$0.2$.
This is in line with the numbers we find in our calculations, for similar parameters statisfying the geometric screening criterion and $Q\lambda_{\rm TF}$.

These crude estimates above clearly indicate the the geometric underscreening is far from negligible for this class of materials, and should be considered seriously even in concrete realistic models.
We stress here once more, that the general conclusions we draw from simplistic models should help drive the search for quantum geometric unconventional superconductors in the future.

\section{Metric transformations}
Start with the line element
\begin{equation}
    ds^2 = g_{xx}dx^2+g_{yy}dy^2+2g_{xy}dxdy.
\end{equation}
Now, use $x=r\cos\theta$, $y=r\sin\theta$.
Consequently,
$dx = dr\cos\theta-rd\theta\sin\theta$,
$dy = dr\sin\theta+rd\theta\cos\theta$.
Hence,
\begin{align}
    ds^2&=dr^{2}\left[g_{xx}\cos^{2}\theta+g_{yy}\sin^{2}\theta+g_{xy}\sin2\theta\right]\nonumber\\
    &+r^{2}d\theta^{2}\left[g_{xx}\sin^{2}\theta+g_{yy}\cos^{2}\theta-g_{xy}\sin2\theta\right]\nonumber\\
    &+rdrd\theta\left[-g_{xx}\sin2\theta+g_{yy}\sin2\theta+2g_{xy}\cos2\theta\right]
\end{align}

In the special case, 
 $g_{xx}=\frac{\zeta^{2}}{4}x^{2}$, $g_{yy}=\frac{\zeta^{2}}{4}y^{2}$, $g_{xy}=\frac{\zeta^{2}}{4}xy$, which corresponds to the small momentum regime in the tunable metric model, one finds the simplified
 \begin{equation}
     ds^2 \approx r^{2}\frac{\zeta^{2}}{4}dr^{2}.
 \end{equation}
 The metric is clearly anisotropic.
 Only translations along the radial direction register on the quantum distance $ds$.

 In another special case, $g_{xx}=g_{yy}\equiv g$, $g_{xy}=0$, corresponding to the topological band inversion model,
 one finds,
 \begin{equation}
     ds^{2}=gdr^{2}+gr^{2}d\theta^{2}.
 \end{equation}
 In the case of the BHZ model, 
 $g=\frac{Q^2}{\left(r^2+Q^2\right)^2}$.
 Defining the variable $\kappa=r/Q$, the line element is simply given by
 \begin{equation}
    ds^{2}=\frac{d\kappa^{2}+\kappa^{2}d\theta^{2}}{\left(\kappa^{2}+1\right)^{2}}.
 \end{equation}
 The numerator shows that the quantum metric is isotropic, the length-squared $ds^2$ does not depend on the direction of an infinitesimal translation.
 However, the denominator clearly indicates that the metric is inhomogenous, and clearly depends on the distance from the origin.

 \section {Contributions from remote bands}

We now wish to explore the influence of the remote bands on the static
polarization, and thus on the Kohn-Luttinger mechanism. 
Here, we show that in scenarios where an isolated band is considered, contributions from remote bands are parametrically smaller as compared to the active-band contribution, with a suppression factor comparable to or smaller than the bandwidth-to-bandgap ratio.
We then perform some explicit calculations for the specific models considered in the main text, confirm this expectation, and reveal other non-universal behaviors.

We denote the dispersion of the $i$ band relative to the Fermi level as $\xi_{{\bf k}}^{i}$,
and the appropriate Bloch wavefunction as $|u_{{\bf k}}^{i}\rangle$.
We denote the active band with the index $i=0$. 
The total static polarization is thus (suppressing the $N_{f}$ prefactor),

\begin{align*}
{\cal P}_{{\bf q}} & =-\sum_{{\bf k},i,j}\left|\left\langle u_{{\bf k+q}}^{i}|u_{{\bf k}}^{j}\right\rangle \right|^{2}\frac{n\left(\xi_{{\bf k+q}}^{i}\right)-n\left(\xi_{{\bf k}}^{j}\right)}{\xi_{{\bf k+q}}^{i}-\xi_{{\bf k}}^{j}}\\
 & =\Pi_{{\bf q}}-\sum_{{\bf k},i\neq0}\left|\left\langle u_{{\bf k+q}}^{i}|u_{{\bf k}}^{0}\right\rangle \right|^{2}\frac{n\left(\xi_{{\bf k+q}}^{i}\right)-n\left(\xi_{{\bf k}}^{0}\right)}{\xi_{{\bf k+q}}^{i}-\xi_{{\bf k}}^{0}}+\sum_{{\bf k},i\in{\rm empty},j\in{\rm full}}\left|\left\langle u_{{\bf k+q}}^{i}|u_{{\bf k}}^{j}\right\rangle \right|^{2}\frac{1}{\xi_{{\bf k+q}}^{i}-\xi_{{\bf k}}^{j}}.
\end{align*}
The first contribution is the active band polarization $\Pi_{{\bf q}}$
which we employ throughout this work. The second contribution is due
to particle-hole fluctuations between remote full bands and the empty
part of the active band, and vice versa. The last part corresponds
to empty-full bands fluctuations. The interplay of the quantum geometric
effects considered in the main text and the contribution from remote
bands can be best analyzed by considering once again the small $q=\left|{\bf q}\right|$
behavior. For simplicity, we employ the rotational invariance symmetry
(which exists explicitly in the models we have analyzed), and define
the generalized inter-band metric by the small $q$ expansion,
\[
\left|\left\langle u_{{\bf k+q}}^{i}|u_{{\bf k}}^{j}\right\rangle \right|^{2}=g^{ij}q^{2}+{\cal O}\left(q^{3}\right).
\]
Notice that as $q\to0$ this overlap tends to zero, since by definition
the wavefunctions of different Bloch bands at the same momentum are
orthogonal. The expansion for ${\cal P}_{{\bf q}}$ is then
\begin{align*}
{\cal P}_{{\bf q}} & \approx\Pi_{0}\left(1-\bar{g}q^{2}\right)\\
 & +q^{2}\left[\sum_{\left|{\bf k}\right|\leq k_{F},i\in{\rm empty}}g^{i0}\frac{1}{\xi_{{\bf k}}^{i}-\xi_{{\bf k}}^{0}}+\sum_{\left|{\bf k}\right|\geq k_{F},j\in{\rm full}}g^{j0}\frac{1}{\xi_{{\bf k}}^{j}-\xi_{{\bf k}}^{0}}\right]\\
 & +q^{2}\sum_{{\bf k},i\in{\rm empty},j\in{\rm full}}g^{ij}\frac{1}{\xi_{{\bf k}}^{i}-\xi_{{\bf k}}^{j}}.
\end{align*}
We can make analytical progress by making some reasonable assumptions.

Assuming all bands have a comparable bandwidth $W$, $\Delta_{ij}$
is the characteristic gap between bands $i,j$, and that the quantum
metric is fairly smooth and its Brilouin zone average for a given
pair of indices is $\bar{g}^{ij}$, we approximate the last line contribution
as
\[
\Pi_{{\rm remote-remote}}\approx\Pi_{0}\left(\sum_{i\in{\rm empty},j\in{\rm full}}\bar{g}^{ij}\frac{W}{\Delta_{ij}}\right)q^{2}.
\]
When the active band is well isolated, $W\ll\Delta_{ij}$ this contribution
is parametrically small compared to the isolated one. For many remote
bands with approximately similar contributions, one may approximate
this contribution by replacing the sum $\sum_{i,j}$ by a factor $\sim\log N_{{\rm bands}}$,
where $N_{{\rm bands}}$ is the number of bands. Notice that if the
remote bands have a larger bandwidth as compared to the isolated one,
as is the usual case when an isolated band is considered, the remote
contribution is made even smaller. We note that this remote-remote
contribution is proportional to the geometry-originated quantum capacitance
of the remote bands, $\sum_{{\bf k},i\in{\rm empty},j\in{\rm full}}\frac{g^{ij}}{\xi_{{\bf k}}^{i}-\xi_{{\bf k}}^{j}}=\frac{e^{2}}{4\pi h}c^{{\rm geo.}}$~\cite{Capacitance_holder_Komissarov2024}.

The contributions arising from remote-active band particle hole fluctuations
may be approximated in a similar form, showing that they are proportional
to the ratio of bandwidth over the band gap. Thus, in the isolated
band limit, the approximation ${\cal P}_{{\bf q}}\approx\Pi_{{\bf q}}$
is well controlled. Let us now proceed to explicitly calculate this
contribution in the models we considered thus far, showing an even
better scaling behavior.

\subsection {The BHZ model}

The wavefunctions of the two different bands for the BHZ model are
\[
|u_{{\bf k}}\rangle=\frac{1}{\sqrt{k^{2}+Q^{2}}}\begin{pmatrix}Q\\
k_{x}+ik_{y}
\end{pmatrix},\,\,\,\,\,|v_{{\bf k}}\rangle=\frac{1}{\sqrt{k^{2}+Q^{2}}}\begin{pmatrix}k_{x}-ik_{y}\\
-Q
\end{pmatrix}.
\]

As such,
\[
\left|\left\langle v_{{\bf k+q}}|u_{{\bf k}}\right\rangle \right|^{2}\approx\frac{Q^{2}}{\left(k^{2}+Q^{2}\right)^{2}}q^{2}+{\cal O}\left(q^{3}\right).
\]
We calculate explicitly (setting a cutoff in momentum space $\Lambda$)
\begin{align*}
\sum_{\left|{\bf k}\right|\geq k_{F}}g^{uv}\frac{1}{\xi_{{\bf k}}^{u}-\xi_{{\bf k}}^{v}} & =\frac{m}{2\pi}\int_{k_{F}}^{\Lambda}kdk\frac{Q^{2}}{\left(k^{2}+Q^{2}\right)^{3}}\\
 & =\Pi_{0}q^{2}\frac{Q^{2}}{\left(k_{F}^{2}+Q^{2}\right)^{2}}\frac{1}{4}\left[1-\left(\frac{\Delta+E_{F}}{\Delta+W}\right)^{2}\right],
\end{align*}
where the bandgap is $\Delta=Q^{2}/\left(2m\right)$, and the bandwidth
is $W=\Lambda^{2}/\left(2m\right)$. Notice that the Fubini-Study
metric of the active $u$ band at the Fermi surface $\bar{g}=\frac{Q^{2}}{\left(k_{F}^{2}+Q^{2}\right)^{2}}$
appears in the above expression. Thus, the full expression for the
static polarization at small momentum transfer may be written as
\[
{\cal P}_{{\bf q}}\approx\Pi_{0}\left[1-\bar{g}\frac{3+\left(\frac{\Delta+E_{F}}{\Delta+W}\right)^{2}}{4}q^{2}\right]=\Pi_{0}\left[1-\bar{g}\frac{3+\left(1-\frac{W-E_{F}}{\Delta+W}\right)^{2}}{4}q^{2}\right].
\]
In the limit $\Delta\gg W$ one recovers the isolated band expression.
Notice, however, that even very far from this limit one recovers at
most a $\sim25\%$ percent correction to the isolated band expression.

\subsection {The tunable metric model}

The wavefunctions of the two bands in the tunable metric model are
\[
|u_{{\bf k}}\rangle=\frac{1}{\sqrt{2}}\begin{pmatrix}e^{i\zeta\frac{\alpha_{{\bf k}}}{2}}\\
ie^{-i\zeta\frac{\alpha_{{\bf k}}}{2}}
\end{pmatrix},\,\,\,\,\,|v_{{\bf k}}\rangle=\frac{1}{\sqrt{2}}\begin{pmatrix}e^{i\zeta\frac{\alpha_{{\bf k}}}{2}}\\
-ie^{-i\zeta\frac{\alpha_{{\bf k}}}{2}}
\end{pmatrix}.
\]
Thus,
\begin{align*}
\left|\left\langle v_{{\bf k+q}}|u_{{\bf k}}\right\rangle \right|^{2} & =\sin^{2}\zeta\frac{\cos\left(k\cos\beta\right)-\cos\left(k\cos\beta+q\right)}{2}\\
 & =\sin^{2}\left[\zeta\sin\left(\frac{q}{2}\right)\sin\left(k\cos\beta+\frac{q}{2}\right)\right]\\
 & \approx\zeta^{2}\frac{q^{2}}{4}\sin^{2}\left(k\cos\beta\right)+{\cal O}\left(q^{3}\right).
\end{align*}
The explicit calculation of the contribution to ${\cal P}_{{\bf q}}$,
\begin{align*}
\sum_{\left|{\bf k}\right|\geq k_{F}}g^{uv}\frac{1}{\xi_{{\bf k}}^{u}-\xi_{{\bf k}}^{v}} & =\zeta^{2}\frac{q^{2}}{4E_{g}}\frac{1}{2\pi}\int_{k_{F}}^{\Lambda}kdk\int\frac{d\beta}{2\pi}\sin^{2}\left(k\cos\beta\right)\\
 & =q^{2}\frac{\zeta^{2}}{8E_{g}}\frac{1}{2\pi}\int_{k_{F}}^{\Lambda}kdk\left[1-J_{0}\left(2k\right)\right]\\
 & \leq\Pi_{0}q^{2}\frac{\left(\zeta a\right)^{2}}{8}\frac{W-E_{F}}{E_{g}},
\end{align*}
where $\Lambda$ is a momentum space cutoff and the bandwidth is $W=\Lambda^{2}/\left(2m\right)$.
The total static polarization may be written as (in the worst case
scenario as far as geometric effect are concerned),
\[
{\cal P_{{\bf q}}}\approx\Pi_{0}\left[1-\bar{g}\left(1-\frac{1}{\left(k_{F}a\right)^{2}}\frac{W-E_{F}}{E_{g}}\right)q^{2}\right].
\]
The remote band contribution is smaller as compared to the purely-active contribution  by roughly the ratio of
the distance between the Fermi level and the top of the band over
the band gap. In the limit $E_{g}\gg W$, appropriate to the isolated
band limit, this is a reasonable approximation. Thus we expect
only small corrections to $\Pi_{{\bf q}}$ in this model due to remote
bands.

\end{widetext}

\end{document}